\documentclass[12pt]{article}
\usepackage{amsmath}               
\usepackage{mathtools}
\usepackage{graphicx,epsf}
\usepackage{enumerate}
\DeclareMathOperator*{\argmin}{argmin}
\DeclareMathOperator*{\argmax}{argmax}
\usepackage{epstopdf}
\usepackage{pdflscape}
\usepackage{graphicx}
\usepackage{subfig}
\usepackage{bm}
\usepackage{tabularx, booktabs}
\usepackage{threeparttable}
\usepackage{ragged2e}
\usepackage{caption}
\usepackage{theorem}
\usepackage{amssymb}
\usepackage{float}
\usepackage[authoryear,round]{natbib}
\usepackage{setspace}
\usepackage{url}
\usepackage{enumitem}
\usepackage[table]{xcolor}
\newcommand{\blind}{0}
\newtheorem{theorem}{Theorem}
\newtheorem{remark}{Remark}
\newtheorem{algorithm}{Algorithm}
\usepackage[page,title]{appendix}
\usepackage{bibentry}
\usepackage{ulem}

\newcommand{\bX}{\boldsymbol{X}}

\newcommand{\bPsi}{\boldsymbol{\Psi}}
\newcommand{\bPhi}{\boldsymbol{\Phi_1}}

\newcommand{\bPi}{\boldsymbol{\Pi}}
\newcommand{\bPit}{\boldsymbol{\Pi^\top}}
\newcommand{\bEta}{\boldsymbol{H}}
\newcommand{\bEtat}{\boldsymbol{H^\top}}
\newcommand{\bG}{\boldsymbol{G}}
\newcommand{\bU}{\boldsymbol{U}}
\newcommand{\bV}{\boldsymbol{V}}
\newcommand{\bx}{\boldsymbol{x}}
\newcommand{\bM}{\boldsymbol{M}}
\newcommand{\bF}{\boldsymbol{F}}
\newcommand{\bu}{\boldsymbol{u}}
\newcommand{\bI}{\boldsymbol{I}}
\graphicspath{{plots/}}

\begin{document}
\def\spacingset#1{\renewcommand{\baselinestretch}%
	{#1}\small\normalsize} \spacingset{1}

\if0\blind	
{
	\title{\bf Homogeneity and Sub-homogeneity Pursuit: Iterative Complement Clustering PCA}
	\author{\\Daning Bi, Le Chang\thanks{Corresponding author, email: le.chang@anu.edu.au}, Yanrong Yang\\
		Research School of Finance, Actuarial Studies \& Statistics\\
			The Australian National University\\}
	\maketitle
} \fi

\if1\blind
{
	\bigskip
	\bigskip
	\bigskip
	\begin{center}
		{\LARGE\bf Homogeneity and Sub-homogeneity Pursuit: Iterative Complement Clustering PCA}
	\end{center}
	\medskip
} \fi

\bigskip
\begin{abstract}
Principal component analysis (PCA), the most popular dimension-reduction technique, has been used to analyze high-dimensional data in many areas. It discovers the homogeneity within the data and creates a reduced feature space to capture as much information as possible from the original data. However, in the presence of a group structure of the data, PCA often fails to identify the group-specific pattern, which is known as sub-homogeneity in this study.  Group-specific information that is missed can result in an unsatisfactory representation of the data from a particular group. It is important to capture both homogeneity and sub-homogeneity in high-dimensional data analysis, but this poses a great challenge. In this study, we propose a novel iterative complement-clustering principal component analysis (CPCA) to iteratively estimate the homogeneity and sub-homogeneity. A principal component regression based clustering method is also introduced to provide reliable information about clusters. Theoretically, this study shows that our proposed clustering approach can correctly identify the cluster membership under certain conditions. The simulation study and real analysis of the stock return data confirm the superior performance of our proposed methods. Supplementary materials, including R codes, for the simulation and real data analysis are available online.
\end{abstract}

\noindent%
\vfill

\newpage
\spacingset{1.5} 
\section{Introduction} \label{intro}

Since its introduction, principal component analysis (PCA) \citep{Jolliffe,anderson2003introduction} has become one of the most popular statistical tools for data analysis in a wide range of areas. Literature on PCA dates back to the early twentieth century (e.g., see \citet{pearson1901liii} and \citet{hotelling1933analysis}). However, with an increasing dimension of data, PCA has now been reconsidered and widely discussed for the purpose of analyzing high-dimensional data.  \citet{jolliffe2016principal} and \citet{fan2018principal} reviewed recent developments in PCA for statistical analysis on high-dimensional data, including its sparsity and robustness. However, are more data together really benefiting statistical analysis? \citet{Boivin2006Are} provided a negative answer at the aspect of forecasting. This is due to an increased complexity when more data from different populations are grouped together, as a proportion of data can exhibit a different pattern compared with the rest. More specifically, increased complexity refers to heterogeneity when more data are collected from different populations.

In this study, we consider that information from a particular group of data collected from one population can be divided into two categories. One is shared with other groups and forms the homogeneity within the entire data, while the other is group-specific and is the main source of heterogeneity. This type of heterogeneity can be treated as sub-homogeneity, which refers to the homogeneity for a particular group of data.

However, sub-homogeneity may not be identified using traditional estimation methods such as PCA. One reason for this is that the sub-homogeneity for a particular group of data can be relatively small compared with the homogeneity within the entire data because it usually contributes less than the homogeneity to the total variance of all of the data. In such a situation, traditional PCA may regard the sub-homogeneity as negligible compared with homogeneity and ignore this group-specific pattern. Moreover, from the interpretation perspective, principal components that are produced using traditional PCA on all of the data do not target a specific group (e.g., information on which component corresponds to which group of data is not known). In previous studies, the discussion of PCA has mainly focused on the large eigenvalues, which correspond to the homogeneity in this study. For example, \citet{johnstone2001distribution} studied the asymptotic distribution of the largest eigenvalue of PCA. Recently, \citet{cai2017limiting} extended the discussion to the asymptotic distribution of the spiked eigenvalues, while \citet{morales2018asymptotics} studied the asymptotics for the leading eigenvalues and eigenvectors of the correlation matrix in the class of spiked models. None of them have considered the existence of sub-homogeneity within the data.

To the best of our knowledge, the sub-homogeneity has not been well discussed, but it can be very important in high-dimensional data analysis. The following example of analyzing stock returns from different industries is used to explain the importance of finding sub-homogeneity. As stated in \citet{fama1997industry}, stock returns from various industries can have varying performance over time, although the homogeneity (e.g., market return) can be deemed to be driven by some common economic variables. In a situation in which a vast number of individual stock returns are collected together for statistical analysis, traditional dimension-reduction techniques such as PCA may be able to capture the market effect but can fail to identify the sub-homogeneity within each industry (industry-specific pattern). This is because some industry-specific components may have much smaller variance than the market component and are highly likely to be omitted by PCA. However, these industry-specific components may be very important in capturing the movement of the stock returns within the industry. This loss of information may result in a very poor forecast of stock returns for some companies in which sub-homogeneity exists. In addition, although it would be interesting to study which industry has a larger industry-specific effect on stock returns, traditional PCA performed on the entire data does not allow us to draw such conclusions. Therefore, this study aims at identifying both homogeneity and sub-homogeneity in high-dimensional data analysis.

The sub-homogeneity in several parts of the data can be identified and estimated by dividing the whole data set into several groups. However, in most situations, the group structure is not known in advance. Therefore, a clustering method can be used to group the data at the first stage, and PCA can then be applied to identify the sub-homogeneity within each group. Similarly, \citet{liu2002block} performed PCA in different blocks of variables, while \citet{tan2015cluster} penalized the variables in each group differently when using the graphical lasso. Both studies used hierarchical clustering to group the variables to take into account the heterogeneity. However, when homogeneity exists across different groups, the sub-homogeneity in each group cannot be successfully identified. This is mainly because the homogeneity shared by most groups can dominate the sub-homogeneity so that the data from different groups all seem to be highly correlated. In this situation, the sub-homogeneity is masked by the homogeneity, and these clustering methods tend to group all individuals (variables) into one cluster. Therefore, the homogeneity must be correctly identified and removed before we can successfully group the individuals and discover the sub-homogeneity of each group.

On the other hand, the estimation of homogeneity can often be problematic in a high-dimensional setting. This is because of the inconsistency of PCA estimates when the number of variables $p$ is comparable with or greater than the sample size $n$, as discussed in \citet{johnstone2009consistency}. Motivated by the group structure of the data, we suggest first clustering the individuals into groups and then performing a traditional approach such as PCA on the level of groups, followed by a second layer of PCA that extracts common information shared by each group. This approach can improve the estimation accuracy of the homogeneity because the components are now extracted from groups in which the dimension has already been reduced (to the number of variables in a group), while the traditional dimension-reduction method (e.g., PCA) is performed on the full dimension $p$. This can be viewed as an effective way to alleviate the potential problem caused by the curse of dimensionality. Other studies have modified the traditional approaches to deal with the curse of dimensionality. For example, \citet{johnstone2009consistency} suggest that some initial dimension-reduction work is necessary before using PCA, as long as a sparse basis exists. Further, \citet{zou2006sparse} introduced sparse PCA, which uses the lasso (elastic net) to modify the principal components so that sparse loadings can be achieved. In addition, a review of some sparse versions of PCA can be found in \cite{tibshirani2015statistical}, while some general discussions about the blessing and curse of dimensionality can be found in \citet{donoho2000high} and \citet{fan2014challenges}.

To conclude, homogeneity must be removed to correctly identify the cluster structure and sub-homogeneity, but the cluster structure of the data is used to accurately find the homogeneity. Therefore, we introduce a novel ``iterative complement-clustering principal component analysis" (CPCA) to iteratively estimate the homogeneity and sub-homogeneity. Details of CPCA are provided in Section \ref{estmation}.

The contributions of this study can be summarized as follows. First, we propose CPCA to identify homogeneity and sub-homogeneity and handle the interaction between them when the whole data set exhibits a group structure. Second, our proposed estimation method not only correctly captures the sub-homogeneity, but also provides very reliable cluster information (e.g., which part of the data is from the same group), which can be useful in understanding and explaining the data structure. Third, inspired by \citet{chiou2007functional}, we develop a leave-one-out principal component regression (PCR) clustering method that can outperform the hierarchical clustering method used in previous studies. In addition, we theoretically illustrate that if the sub-homogeneity from different clusters is distinct, our proposed clustering procedure can effectively separate the variables from different clusters. This is also numerically confirmed by the simulation study and real data analysis. Details of the proposed clustering method are provided in Section \ref{estmation}.

The rest of this study is organized as follows. A low-rank representation of the data that captures both homogeneity and sub-homogeneity is introduced in Section \ref{model}, and some related PCA methods are discussed in Section \ref{other model}. In Section \ref{estmation}, a novel estimation method called CPCA is proposed, followed by a discussion of more details of the algorithm. Section \ref{cluster theory} demonstrates and explains the effectiveness of the proposed clustering method. Extensive simulations, along with two applications (PCR and covariance estimation) of our proposed method are provided in Sections \ref{sim} and \ref{application}. Section \ref{real} analyzes a stock return dataset using our method. Lastly, Section \ref{con} concludes the study.

\section{Homogeneity and Sub-homogeneity} \label{model}
Considering the data $\boldsymbol{x}_i=(x_{1i}, \ldots,x_{pi})^\top \in \mathbb{R}^p$ from $i^{th}$ observation with $p$ variables, the singular value decomposition (SVD) of the data $\boldsymbol{x}_i$ can be written as:
\begin{align}
	\nonumber\boldsymbol{x}_i =& \sum_{k=1}^{K}{g}_{ik}\boldsymbol{\phi}_k +\boldsymbol{u}_{i},\\ \text{where } \boldsymbol{u}_{i}=&\sum_{k=K+1}^{\min{(n,p)}}{g}_{ik}\boldsymbol{\phi}_k\quad i=1,\ldots, n,
	\label{svd}
\end{align}
where ${g}_{ik}$ is defined as $k^{th}$ principal component score and $\boldsymbol{\phi}_k$ denotes a $p\times{1}$ eigenvector for ${g}_{ik}$. Traditional PCA summarizes the data using the first $k$ principal components, and it treats $\boldsymbol{u}_{i}$ as noise because ${g}_{ik}$ for $k=m+1,\ldots,\min{(n,p)}$ has smaller variance. However, under certain conditions, some of the information contained in $\boldsymbol{u}_{i}$ can be useful in prediction or forecasting problems, particularly for one or more specific groups of data. Therefore, when the data exhibit a group structure, we propose the following low-rank representation for the data to capture both homogeneity and sub-homogeneity:
\begin{align}
	\nonumber\boldsymbol{x}_i = &\sum_{k=1}^{r_c}{g}_{ik}\boldsymbol{\phi}_k +\boldsymbol{u}_{i},\\ \text{where }
	\boldsymbol{u}_{i}=\sum_{j=1}^{J}\sum_{h=1}^{r_j}{f}^{(j)}_{ih}\boldsymbol{\gamma}^{(j)}_h& + \sum_{j=1}^{J}\boldsymbol{I}^{(j)}\boldsymbol{\epsilon}^{(j)}_{i}, \quad i=1,\ldots, n, \quad j=1,\ldots, J.
	\label{model1}
\end{align}
The first line in (\ref{model1}) measures the homogeneity among all variables from all groups, where ${g}_{ik}, k=1,\ldots, r_c$ is  $k^{th}$ principal component, which we call the common component, and $\boldsymbol{\phi}_k$ is its corresponding eigenvector. The first part of $\boldsymbol{u}_{i}$ in (\ref{model1}) consists of $J$ cluster-specific components from which the sub-homogeneity of the data is derived. Assuming $p$ variables can be split into $J$ clusters, ${f}^{(j)}_{ih}, h=1,\ldots,r_j$ measures the within-cluster principal components for cluster $j$. The within-cluster eigenvector with dimension $p \times 1$ has the form of $\boldsymbol{\gamma}^{(j)}_h=(\boldsymbol{0}^{(1)\top},\ldots,\boldsymbol{\eta}^{(j)\top}_h,\ldots,\boldsymbol{0}^{(J)\top})^\top$, where $\boldsymbol{\eta}^{(j)}_h$ defines a $p_j\times1$ vector and $p_j$ is the number of variables in cluster $j$ so that $\sum_{j=1}^{J}p_j=p$. That is, the values of $\boldsymbol{\gamma}^{(j)}_h$ for variables that do not belong to cluster $j$ are zero. This implies that after removing the effect of the common principal components, variables from different clusters are uncorrelated (e.g., $\boldsymbol{u}_{i}$ has a block-diagonal covariance structure). In the second part of $\boldsymbol{u}_{i}$, $\boldsymbol{\epsilon}^{(j)}_{i}$ is simply a $p_j$-dimensional error that has variance ${\sigma^{(j)}}^2$, and $\boldsymbol{I}^{(j)}$ is a diagonal matrix, but the diagonals for variables that do not belong to cluster $j$ are zero.

We further define $\boldsymbol{g}_i=({g}_{i1},\ldots,{g}_{ir_c})^\top$ and $\boldsymbol{f}^{(j)}_i=({f}^{(j)}_{i1},\ldots,{f}^{(j)}_{ir_j})^\top, j=1,\ldots,J$ as the vector forms of the common components and cluster-specific components, and the eigenvector matrices as $\boldsymbol{\Phi}_{(p\times r_c)}=(\boldsymbol{\phi}_1,\ldots,\boldsymbol{\phi}_{r_c})$ and $\boldsymbol{\Gamma}^{(j)}_{(p\times r_j)}=(\boldsymbol{\gamma}^{(j)}_1,\ldots,\boldsymbol{\gamma}^{(j)}_{r_j}), j=1,\ldots,J$,  respectively. Without loss of generality, we assume $E(\boldsymbol{g}_i)=E(\boldsymbol{f}^{(j)}_i)=E(\boldsymbol{\epsilon}^{(j)}_{i})=\boldsymbol{0}$, and $\boldsymbol{g}_i,\boldsymbol{f}^{(j)}_i,\boldsymbol{\epsilon}^{(j)}_{i}, j=1,\ldots, J$ are mutually uncorrelated. In addition, we also impose the following canonical condition for the identification of both the homogeneity and sub-homogeneity,
\begin{align}
	\nonumber
	 \boldsymbol{\Phi}^\top\boldsymbol{\Phi}=I_{r_c},& \quad  \boldsymbol{\Sigma}_g   \coloneqq     \text{cov}(\boldsymbol{g}_i) \text{ is diagonal},\\  
	 {\boldsymbol{\Gamma}^{(j)\top}}\boldsymbol{\Gamma}^{(j)} =I_{r_j},& \quad 
	 \boldsymbol{\Sigma_}f^{(j)}  \coloneqq   \text{cov}(\boldsymbol{f}^{(j)}_i)
	\text{ is diagonal}, \quad j=1,\ldots, J.
	\label{ident}
\end{align}
Under the data structure given in (\ref{model1}), the population covariance matrix is given by a low-rank plus block-diagonal representation
\begin{align}
	\boldsymbol{\Sigma}= \boldsymbol{\Phi}\boldsymbol{\Sigma}_g \boldsymbol{\Phi}^\top+ \sum_{j=1}^{J}\boldsymbol{\Gamma}^{(j)}\boldsymbol{\Sigma}_f^{(j)} {\boldsymbol{\Gamma}^{(j)\top}}+ \sum_{j=1}^{J}\boldsymbol{I}^{(j)}{\sigma^{(j)2}}, \quad j=1,\ldots, J.
	\label{cov}
\end{align}

If we denote the data $\boldsymbol{X}_{(n\times p)}=(\boldsymbol{x}_1,\ldots,\boldsymbol{x}_n)^\top$, the common components and cluster-specific components in a matrix form $\boldsymbol{G}_{(n\times{r_c})}=(\boldsymbol{g}_1,\ldots,\boldsymbol{g}_n)^\top$ and $\boldsymbol{F}^{(j)}_{(n\times{r_j})}=(\boldsymbol{f}^{(j)}_1,\ldots,\boldsymbol{f}^{(j)}_n)^\top$ , respectively, data representation (\ref{model1}) can also be presented in the following matrix form
\begin{align}
	\nonumber\boldsymbol{X} =& \boldsymbol{G} {\boldsymbol{\Phi}}^\top+\boldsymbol{U},\\ \text{where } \boldsymbol{U}=&\sum_{j=1}^{J}\boldsymbol{F}^{(j)}{\boldsymbol{\Gamma}^{(j)\top}} + \boldsymbol{E}, \quad j=1,\ldots, J.
	\label{model3}
\end{align}

In general, there are two goals in using representation (\ref{model1}). In the presence of an unknown cluster structure, the first goal is to cluster the variables (i.e., determine which variables are in the same group). From the interpretation perspective, it is interesting to know how variables are clustered and which variables belong to the same cluster. The second goal is to correctly estimate both the common components $\boldsymbol{g}_i$ and the cluster-specific components $\boldsymbol{f}^{(j)}_i$. These components serve as a low-rank representation of the data and can therefore be used for further applications. One obvious application is to estimate the covariance matrix $\boldsymbol{\Sigma}$ and the precision matrix $\boldsymbol{\Sigma}^{-1}$ of $\boldsymbol{X}$. The use of both $\boldsymbol{g}_i$ and $\boldsymbol{f}^{(j)}_i$ captures both the low rank and the block-diagonal representation of $\boldsymbol{\Sigma}$, which results in a more efficient estimation compared with using $\boldsymbol{g}_i$ only. Another important application is PCR. In some situations, it is a cluster-specific component $\boldsymbol{f}^{(j)}_i$ that contributes most to determining the response in PCR rather than the common component $\boldsymbol{g}_i$. It is important to identify each cluster-specific component $\boldsymbol{f}^{(j)}_i$ to explain which cluster of variables has a greater effect in predicting the response. We will discuss more about these two applications using simulated data in Section \ref{application}.

\section{Relationship with Existing PCA Methods}\label{other model}

Our proposed data representation (\ref{model1}) should be used to perform dimension reduction of the data with a cluster structure because it captures both the common effect (homogeneity) and the cluster-specific effect (sub-homogeneity). Consequently, this method can be viewed as an extension of many other widely used dimension-reduction methods. Three of these methods are discussed below.

\begin{itemize}
	\item \textsc{Case} 1: When there is no cluster-specific effect, for example, $\boldsymbol{f}^{(j)}_i=\boldsymbol{0}, j=1,\ldots, J$ and ${\sigma^{(j)2}}\equiv{\sigma}^2$, representation (\ref{model1}) simply reduces to the well-known PCA
	\begin{align}
		\boldsymbol{x}_i = \sum_{k=1}^{r_c}{g}_{ik}\boldsymbol{\phi}_k +\boldsymbol{\epsilon}_{i}, \quad i=1,\ldots, n,
		\label{case1}
	\end{align}
	where  $\boldsymbol{g}_{ik}$ can be found by the principal component with $k^{th}$ largest eigenvalue and $\boldsymbol{\phi}_k $ is the corresponding eigenvector. However, in the presence of a small cluster-specific effect, traditional PCA generally only identifies the large common effect and treats the rest as noise, disregarding the sub-homogeneity within the data.
	
	\item \textsc{Case} 2: When the common components do not exist (e.g., $\boldsymbol{g}_i=\boldsymbol{0}$), representation (\ref{model1}) demonstrates a well-studied block-diagonal covariance structure of the data, as discussed in \citet{liu2002block} and \citet{tan2015cluster}. Thus, representation (\ref{model1}) can be reduced to
	\begin{align}
		\boldsymbol{x}_i = \sum_{j=1}^{J}\sum_{h=1}^{r_j}{f}^{(j)}_{ih}\boldsymbol{\gamma}^{(j)}_h + \sum_{j=1}^{J}\boldsymbol{I}^{(j)}\boldsymbol{\epsilon}^{(j)}_{i}, \quad i=1,\ldots, n, \quad j=1,\ldots, J.
		\label{case 2}
	\end{align}
	In this case, hierarchical clustering is usually used to group the variables with high correlations, and PCA is then performed on each group of variables to estimate the cluster-specific components $\boldsymbol{f}^{(j)}_i$. In many situations, this cluster structure is masked by a dominant common effect. Ignoring this common effect will result in a non-identifiable cluster structure.
	
	\item \textsc{Case} 3: \citet{fan2013large} and \citet{li2018embracing} consider the following low-rank representation, in which the error covariance matrix is assumed to be cross-sectional dependent after the common components have been taken out and
	\begin{align}
		\boldsymbol{x}_i = \sum_{k=1}^{r_c}{g}_{ik}\boldsymbol{\phi}_k +\boldsymbol{u}_{i}, \quad i=1,\ldots, n,
		\label{case3}
	\end{align}
	where they assume that the covariance of $\boldsymbol{u}_{i}$, $\Sigma_u$, is sparse and propose a method called Principal Orthogonal complEment Thresholding (POET) to explore such a high-dimensional structure with sparsity. \citet{li2018embracing} used weighted PCA to find a more efficient estimator of common components $\boldsymbol{g}_i$. \citet{hong2018optimally} and \citet{deville1983data} also applied weighted PCA to estimate the covariance matrix when heteroscedastic noise of samples or variables exists. However, in this study, we consider $\boldsymbol{u}_{i}$ following a cluster structure with a block-diagonal covariance and aim to find its low-rank presentation, which we call sub-homogeneity. In addition, we propose iteratively estimating the common components and cluster-specific components.
\end{itemize}

\section{Estimation Methods}\label{estmation}

\begin{figure}[!htbp]
	\centering
	\includegraphics[scale=1]{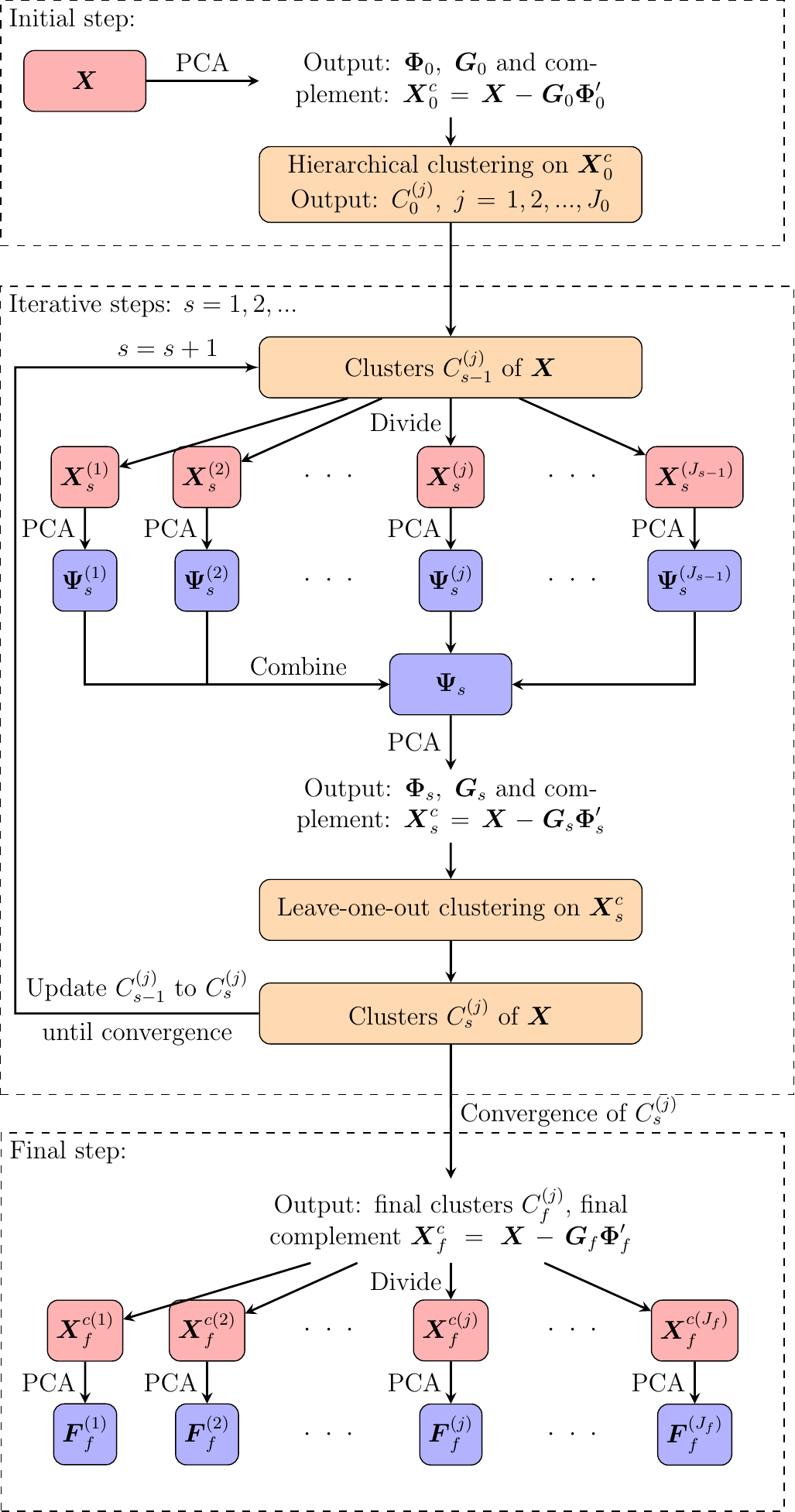}
	\caption{Flowchart of Algorithm 1.}
	\label{fig:flowchart} 
\end{figure}

Correctly estimating $\boldsymbol{g}_i$ and $\boldsymbol{f}^{(j)}_i$ poses many challenges. First, PCA, which is widely used to estimate the common component $\boldsymbol{g}_i$, often performs very poorly when $p$ is much larger than $n$. The group structure of the variables motivates us to separate the data according to the clusters and then extract the common information from each cluster to determine the common components. This is less influenced by the curse of dimensionality because each cluster of the data results in a lower dimension of variable $p_j$. However, to accurately identify the clusters, we must remove the common effect and then perform the clustering method based on the complement $\left(\boldsymbol{x}_i- \sum_{k=1}^{r_c}{g}_{ik}\boldsymbol{\phi}_k\right)$, but the common components and its eigenvectors are not known in advance. This inspires us to propose a novel iterative method, CPCA, to cluster the variables and estimate the components simultaneously. The details of the method are described in Algorithm 1 and a flowchart that summarizes our estimation method is presented in Figure \ref{fig:flowchart}.

\begin{algorithm}[CPCA]
	\label{algorithm} 
	\item              
	\begin{enumerate}
		\item Initial Step:
		\begin{enumerate}[label=(\alph*)]
			\item Perform PCA directly on the entire data $\boldsymbol{X}$ and select the number of components using a ratio based estimator defined by \citet{lam_factor_2012}. A brief discussion on selecting the number of components is in Remark \ref{remark1}. The resulting eigenvectors $\boldsymbol{\Phi}_0$ and principal components $\boldsymbol{G}_0$ are served as the initial estimates of $\boldsymbol{\Phi}$ and $\boldsymbol{G}$. Then, find the initial complement $\boldsymbol{X}^c_0=\boldsymbol{X}-	\boldsymbol{G}_0\boldsymbol{\Phi}^\top_0$ 
			
			\item Perform hierarchical clustering for $\boldsymbol{X}^c_0$ based on a similarity matrix given by the absolute value of the empirical correlation matrix $\boldsymbol{X}^c_0$. The obtained clusters $C^{(j)}_0, j=1,\ldots,J_0$ are served as the initial clusters.
		\end{enumerate}
		\item Iterative Step: for $s=1,2,\ldots$
		\begin{enumerate}[label=(\alph*)]
			\item Cluster the variables into $J_{s-1}$ groups according to $C^{(j)}_{s-1}$ and define variables from $j^{th}$ cluster as  $\boldsymbol{X}^{(j)}_s$. Perform PCA on each cluster of variables $\boldsymbol{X}^{(j)}_s$ and denote the obtained principal components as $\boldsymbol{\Psi}^{(j)}_s, j=1,\ldots,J_{s-1}$. Combine these principal components as $\boldsymbol{\Psi}_s=(\boldsymbol{\Psi}^{(1)}_s,\ldots,\boldsymbol{\Psi}^{(J_0)}_s)$ and perform a further step of PCA on $\boldsymbol{\Psi}_s$. Define the  principal components as $\boldsymbol{G}_s$ and their corresponding eigenvectors as $\boldsymbol{\Phi}_s$. Then, compute the updated complement $\boldsymbol{X}^c_s=\boldsymbol{X}-\boldsymbol{G}_s\boldsymbol{\Phi}^\top_s$. Details for finding the eigenvectors $\boldsymbol{\Phi}_1$ can be found in Appendix \ref{append A}.
			
			\item Perform leave-one-out clustering for variables in $\boldsymbol{X}^c_s$ using PCR (more details of this clustering method are discussed in Remark 3):
			\begin{enumerate}[label=(\roman*)]
				\item For $k, k=1,\ldots,p$, leave $k^{th}$ variable out of $\boldsymbol{X}^c_s$.
				\item Group the rest of the variables of $\boldsymbol{X}^c_s$ based on $C^{(j)}_{s-1}$, perform PCA on each cluster again and denote the obtained components as $\boldsymbol{F}_{s-1}^{(j)}, j=1,\ldots,J_{s-1}$.
				\item Fit $J_{s-1}$ PCRs by using $k^{th}$ variable of $\boldsymbol{X}^c_s$ as the response and $\boldsymbol{F}_{s-1}^{(j)}$ as the predictor for each $j=1,\ldots,J_{s-1}$, respectively. 
				\item Compute the sum of squared residuals (SSR) for each $J_{s-1}$ PCR model. Assign $k^{th}$ variable to the cluster with the minimum SSR. Update the cluster index for $k^{th}$ variable in $C^{(j)}_{s-1}$.
				\item Repeat (i)$\sim$(iv) for each $k$ and denote the updated clusters as $C^{(j)}_s$. 
			\end{enumerate}	
			
			\item Repeat (a) and (b) within this step until the clusters converge, and define this final converged cluster as $C^{(j)}_f$.
		\end{enumerate}
		\item Final Step:
		\begin{enumerate}[label=(\alph*)]
			\item Repeat (a) in the Iterative Step, but using cluster $C^{(j)}_f$. The final complement is denoted by $\boldsymbol{X}^c_f=\boldsymbol{X}-\boldsymbol{G}_f\boldsymbol{\Phi}^\top_f$.
			\item Cluster variables of $\boldsymbol{X}^c_f$ based on $C^{(j)}_f$, perform PCA on each cluster $\boldsymbol{X}^{c{(j)}}_f$ and denote the obtained cluster-specific components as $\boldsymbol{F}_f^{(j)}, j=1,\ldots,J_f$.
		\end{enumerate}
	\end{enumerate}
\end{algorithm}

Therefore, CPCA produces these required outputs: the final clusters of $p$ variables $C^{(j)}_f, j=1,\ldots,J_f$, the final estimate of the common components $\boldsymbol{G}_f$ and the cluster-specific components $\boldsymbol{F}_f^{(j)}, j=1,\ldots,J_f$, along with their eigenvectors. Some details and discussions of CPCA are provided in the following remarks.

\begin{remark}
	\label{remark1} 
	\normalfont \textbf{[Selecting the Number of Components]}:\\
	In this work, we use a ratio based estimator \citep{lam_factor_2012} to select the number of common components and estimate the homogeneity. The ratio based estimator of the number of principal components $r_c$ is defined as
	\begin{align}\label{rratio}
		\widehat{r}_c = \argmin_{1\le i \le R}  \frac{\widehat{\lambda}_i}{\widehat{\lambda}_{i+1}},
	\end{align}
	where $\widehat{\lambda}_1 \ge \cdots \ge \widehat{\lambda}_p$ are eigenvalues of the sample covariance matrix of data $\boldsymbol{X}$ and $r \le R \le p$ is a constant. In practice, many researchers may prefer to select the number of principal components by either setting a threshold on the proportion of total variance explained, which may accidentally include some errors into the principal components, or choosing the numbers where the largest drop of eigenvalues exists (i.e. based on the scree plot). Nonetheless, when the number of common components is greater than 1 (e.g., $r_c>1$), the discrepancy of the eigenvalues (from the sample covariance) among the common components can be accidentally greater than the one between the common components and the cluster-specific component, especially when some groups have a relatively larger cluster-specific effect (e.g., variables in one group are on a large scale or highly correlated). In this situation, selecting the number of components using the largest drop in eigenvalues in Initial Step (a) and Iterative Step (a) may lead to underestimating the number of common components, while the ratio-based estimator $\widehat{r}_c$ is less affected by the scales of variables.
	In addition, to consider a broader class of sub-homogeneities, we also allow for sub-homogeneities within a specific cluster to have different order of variances. To correctly estimate the number of principal components that may have different orders of variances in the same group, in Iterative Step 2 (a) and (b), we use the iterative estimation procedure purposed by \citet{peng2020interactive}, which is a generalization of the original ratio-based estimator. The exact procedure is lengthy and is not the focus of the present work, hence is omitted here. More details of the iterative estimation procedure of the ratio-based estimator can be found in Section 2 of \citet{peng2020interactive}. 
\end{remark}

\begin{figure}[!htbp]
	\centering
	
	\subfloat[][]{\includegraphics[scale=0.19]{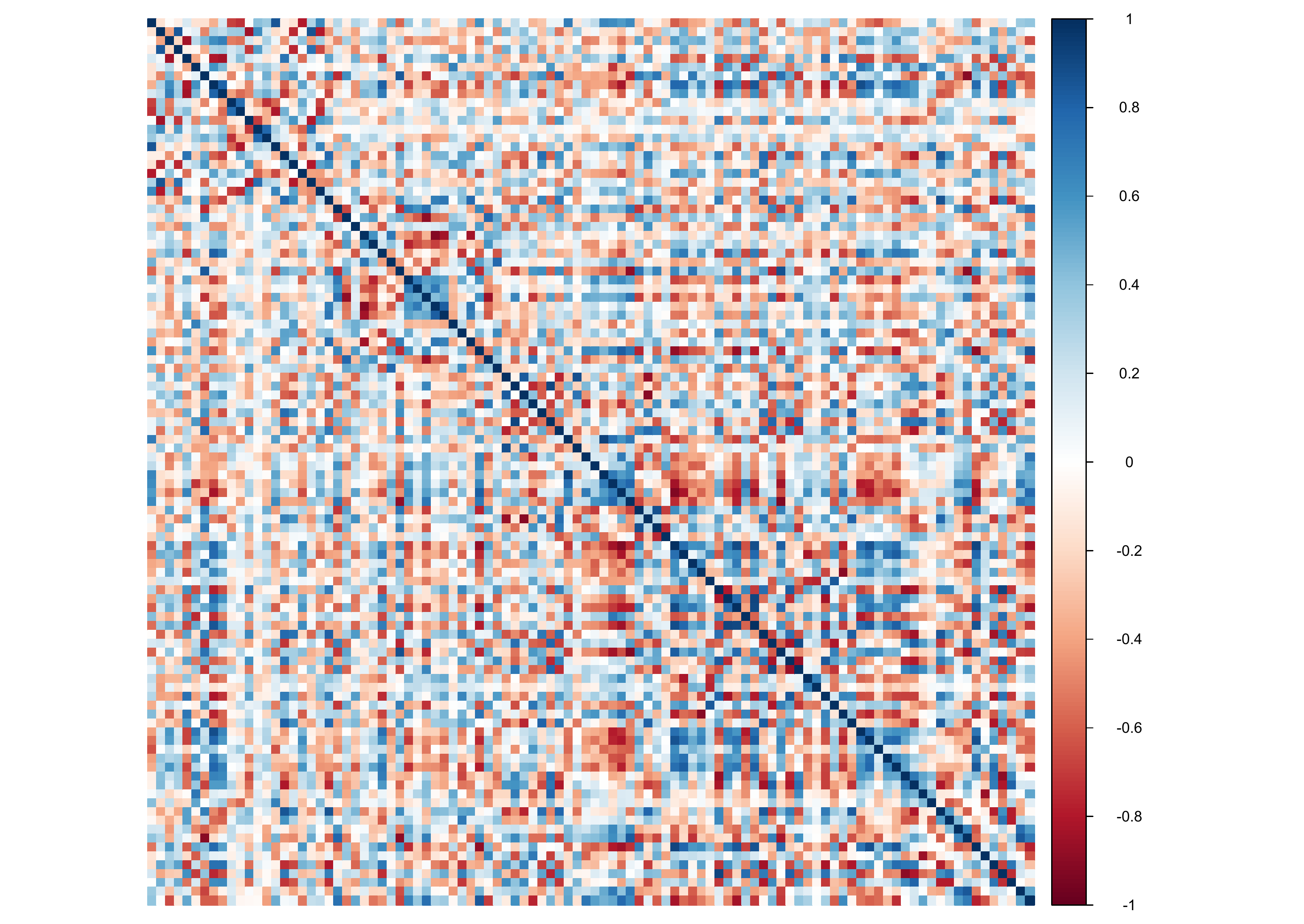}}\subfloat[][]{\includegraphics[scale=0.19]{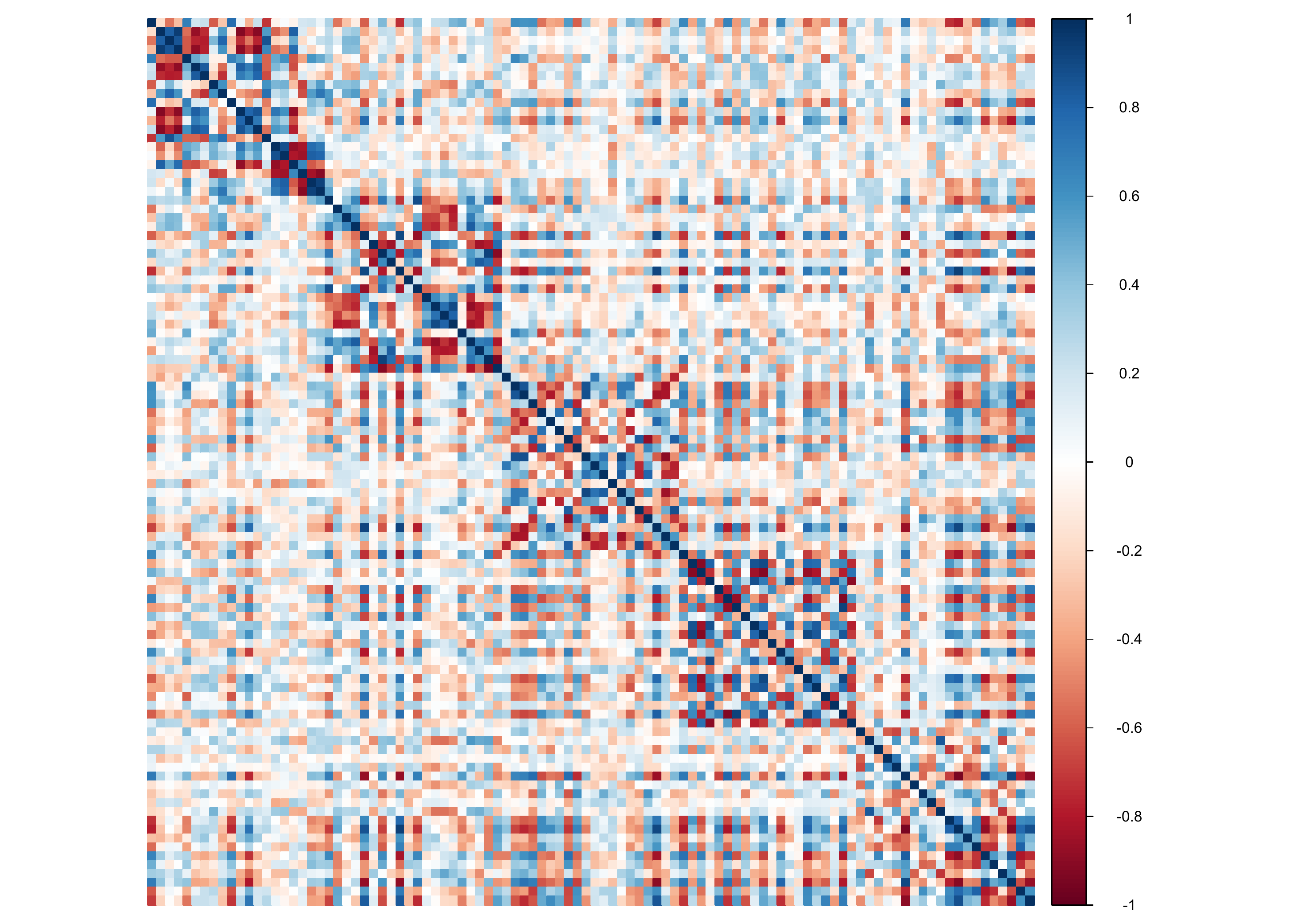}}\subfloat[][]{\includegraphics[scale=0.19]{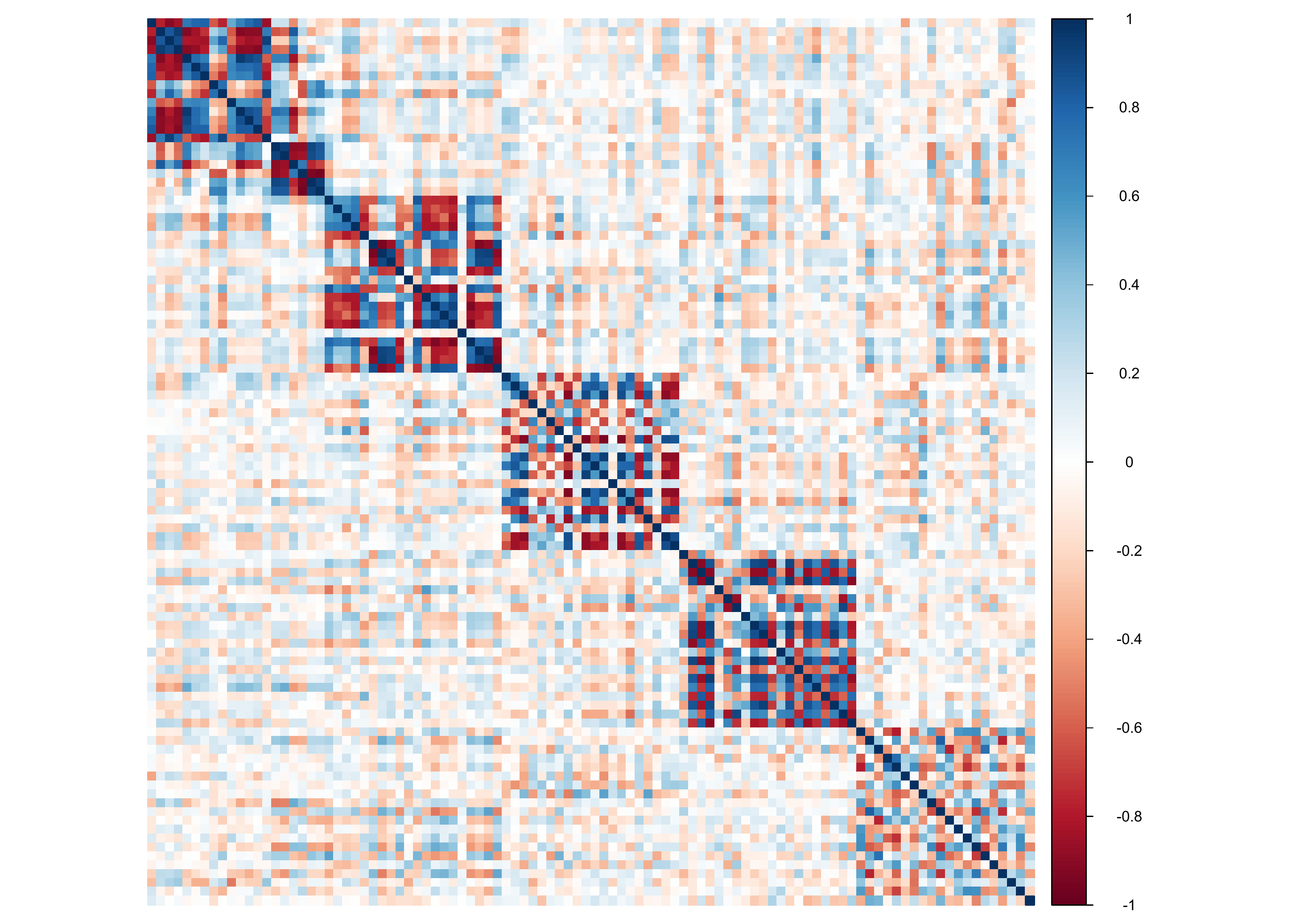}}\\ [10pt]
	
	\caption{Correlation plot using data generated by simulation example 2 for (a) the original data $\boldsymbol{X}$, (b) the initial complement $\boldsymbol{X}^c_0=\boldsymbol{X}-	\boldsymbol{G}_0\boldsymbol{\Phi}^\top_0$ and (c)  the final complement $\boldsymbol{X}^c_f=\boldsymbol{X}-\boldsymbol{G}_f\boldsymbol{\Phi}^\top_f$}
	
	\label{fig:corplot}
\end{figure}

\begin{remark}
	\label{remark2} 
	\normalfont \textbf{[Iterative Step]}:\\
	One of the key contributions of our algorithm is iteratively estimating the common components and cluster-specific components. Directly applying PCA on the entire data $\boldsymbol{X}$ generally leads to a very poor estimate of the common components because a large $p$ may blur the spike structure of the sample covariance matrix. Motivated by the group structure of the data, we utilize Iterative Step (a) to first cluster variables and then perform PCA on the level of clusters, followed by a second layer of PCA that extracts common information shared by each group. 
	Iterative Step (b) is then implemented to update and improve the clusters information. This is more effective in estimating the common components because PCA is performed in a smaller $p$ case, which is also numerically shown in Figure \ref{fig:corplot}. The data $\boldsymbol{X}$ used here are generated by simulation Example 2. Figure \ref{fig:corplot} (a) presents the correlation plot for the original data $\boldsymbol{X}$. We observe that the cluster structure is masked by the common effect. Figure \ref{fig:corplot} (b) demonstrates that after the common effects estimated in the Initial Step are removed, it is still difficult to perceive the cluster structure. However, if we remove the common effects estimated in the Final Step, the block-diagonal structure is clear, implying a prominent heterogeneity within the data. Therefore, iteratively estimating the common components is advantageous.
	
\end{remark}

\begin{remark}
	\label{remark3} 
	\normalfont \textbf{[Initial Step (b), Iterative Step (b) and (c)]}:\\
	Initial Step (b), an ordinary hierarchical clustering method using average-linkage as dissimilarity between clusters is performed. As discussed in \citet{buhlmann2013correlated}, we choose the number of clusters using the largest increment in height between iterations proceeded in a agglomerative way, such that $J_0=\argmax_j(h_{j+1}-h_j)$. 
	
	In Iterative Step (b), we propose a new clustering method called ``leave-one-out PCR clustering" (LOO-PCR clustering), which is another key contribution of CPCA. The underlying idea of this new clustering method is that if one variable belongs to a cluster, it should be well-predicted by the principal components extracted from that cluster. This is more aligned with our model set-up than hierarchical clustering based on correlations. A similar idea of leaving one out is used for functional clustering, as discussed in \citep{chiou2007functional}. In addition, hierarchical clustering can perform poorly when the dimension of variables $p$ is larger than $n$. In step (iv), we assign $k^{th}$ variable to the cluster that achieves the minimum SSR. However, when the minimum SSR is larger than a threshold $\tau$, we treat the $k^{th}$ variable itself as a cluster, because in such a situation, this variable cannot be well-predicted by any clusters of variables. Hence, the number of clusters is driven by the data and can vary in each iteration. This suggests that our proposed method is more flexible than hierarchical clustering. We set $\tau=0.95$, and Example 4 in the simulation studies demonstrates that our proposed LOO-PCR clustering, served as a clustering method itself, outperforms hierarchical clustering in a large $p$ small $n$ case. Identifiability of LOO-PCR clustering is also demonstrated theoretically in Section \ref{cluster theory}.
	
	In Iterative Step (c), we stop the iteration when the clusters $C^{(j)}_s$ converge. In our algorithm, we adopt the adjusted rand index (ARI) \citep{rand1971objective,hubert1985comparing} between the clusters $C^{(j)}_s$ and the one in the previous iteration $C^{(j)}_{s-1}$ as the stopping criterion. The ARI is a corrected version of the rand index that measures the similarity between two clusters of the same data using the proportions of agreements between these two partitions. The correction is achieved by subtracting the rand index by its expected value. The ARI can yield a maximum value of 1, and a high value implies high similarity. When the ARI is above a certain threshold $\eta$, we stop the iteration. In this study, we use $\eta=0.95$.

\end{remark}

\section{Identifiability of LOO-PCR Clustering Approach}\label{cluster theory}
Achieving an accurate sub-homogeneity pursuit relies on the effectiveness of the clustering procedure. In this section, we investigate the proposed LOO-PCR clustering method theoretically. More specifically, we examine the identifiability of cluster membership in our proposed method.

Consider one random variable $X_{mi}$ that belongs to cluster $l$. Based on the variable’s structure in cluster $l$, there exists $m\in\{1, 2, \ldots, p\}$, such that:
\begin{eqnarray}\label{y0}
X_{mi}=\sum^{r_c}_{k=1}g_{ik}\phi_{mk}+u_{mi}^{(l)}, \ \ 
u_{mi}^{(l)}=\sum^{r_l}_{h=1}f_{ih}^{(l)}\gamma_{mh}^{(l)}+\epsilon_{mi}^{(l)}. 
\end{eqnarray}
We now consider another cluster $d$ with components $g_{ik}, k=1, \ldots, r_c$ and $f_{ih}^{(d)}, h=1, \ldots, r_d$. When applying our proposed LOO-PCR clustering method, we regress $u_{mi}^{(l)}$ on $f_{ih}^{(d)}, h=1, 2, \ldots, r_d$ and measure its goodness of fit to determine whether the random variable $X_{mi}$ belongs to cluster $d$, such that:
\begin{eqnarray}\label{y1}
u_{mi}^{(l)}=\sum^{r_d}_{h=1}\beta_hf_{ih}^{(d)}+\zeta_{mi}^{(d)}, 
\end{eqnarray}
where $\beta_1, \ldots, \beta_{r_d}$ are coefficients and $\zeta_{mi}^{(d)}$ is the error. Naturally, we expect that features (e.g., principal components) from cluster $d$ cannot explain $u_{mi}^{(l)}$ sufficiently. Based on (\ref{y0}) and (\ref{y1}), intuitively, a large discrepancy between principal components $f_{ih}^{(d)}$ from cluster $d$ and principal components $f_{ih}^{(l)}$ from cluster $l$ will result in a large residual $\zeta_{mi}^{(d)}$ in (\ref{y1}).

The following theorem will show the property of $\widehat{\zeta}_{mi}^{(d)}$, which is an estimator of the error $\zeta_{mi}^{(d)}$ from our proposed clustering approach.

\begin{theorem}\label{thm1}
	For any cluster $d$ and any random variable $X_{mi}$ from another cluster $l$, we have the following evaluation:
	\begin{eqnarray}\label{yr2}
	\|\widehat{\boldsymbol{\zeta}}_{m}^{(d)}\|_2
	=\|\bM_{\bF^{(d)}}\bM_{\bG}\bx_{m}\|_2+O_p\left(\max(\alpha_{np_d}, \gamma_{nJ})\right)\|\bx_{m}\|_2,  
	\end{eqnarray}
	where $\|\bM_{\widehat{\bF}^{(d)}}-\bM_{\bF^{(d)}}\|_2:=O_p(\alpha_{np_d})$, $\|\bM_{\widehat{\bG}}-\bM_{\bG}\|_2:=O_p(\gamma_{nJ})$, $\widehat{\boldsymbol{\zeta}}_{m}^{(d)}=\left(\widehat{\zeta}_{m1}^{(d)}, \widehat{\zeta}_{m2}^{(d)}, \ldots, \widehat{\zeta}_{mn}^{(d)}\right)^{\top}$, $\bx_{m}=(X_{m1}, X_{m2}, \ldots, X_{mn})^{\top}$, $\bM_{\bF^{(d)}}=\bI_n-\bF^{(d)}\left(\bF^{(d)\top}\bF^{(d)}\right)^{-1}\bF^{(d)\top}$ and $\bM_{\bG}=\bI_n-\bG(\bG^{\top}\bG)^{-1}\bG^{\top}$; here, $\bF^{(d)}$ is an $n\times r_d$ matrix with $(i, h)$-th element being $f_{ih}^{(d)}$, and $\bG$ is an $n\times r_c$ matrix with $(i, k)$-th element being $g_{ik}$. Here, $\bM_{\widehat{\bF}^{(d)}}$ and $\bM_{\widehat{G}}$ are $\bM_{\bF^{(d)}}$ and $\bM_{\bG}$, but with $\bF^{(d)}$ and $\bG$ replaced by $\widehat{\bF}^{(d)}$ and $\widehat{\bG}$, respectively.
\end{theorem}
A brief proof of Theorem \ref{thm1} is provided in Appendix \ref{append B}.
\begin{remark}
	\label{remark5} 
	\normalfont It is expected that $\widehat{\boldsymbol{\zeta}}^{(d)}_{m}$ will be large when cluster $d$ and cluster $l$ are different. When $l=d$, the first term on the right-hand side of (\ref{yr2}) is equal to $\|\bM_{\bF^{(d)}}\bM_{\bG}{\boldsymbol{\zeta}}_{m}^{(d)}\|_2$.  In contrast, when $l\neq d$ (e.g., $\|{\bF^{(l)}}\|_2$ and $\|{\bF^{(d)}}\|_2$ are distinct), the first term on the right-hand side of (\ref{yr2}) is equal to $\|\bM_{\bF^{(d)}}\bM_{\bG}{\bF^{(l)}}{\boldsymbol{\gamma}^{(l)}}+\bM_{\bF^{(d)}}\bM_{\bG}{\boldsymbol{\zeta}}_{m}^{(l)}\|_2$, which is dominated by $\|{\bF^{(l)}}\|_2$. Meanwhile, the second term on the right-hand side of (\ref{yr2}) is determined by the estimation accuracy of $\bF^{(d)}$ and $\bG$. It is related to the dimension of cluster $d$ (e.g., $p_d$), the sample size $n$ and the total number of clusters $J$. \cite{bai2002determining} provided the rate of convergence for the projection matrices of principal components. \cite{fan2013large} and \cite{fan2018principal} also studied the properties of principal components for high-dimensional data. In this study, we do not pursue the exact expression of $\alpha_{np_d}$ and $\gamma_{nJ}$. However, given that the homogeneity  $\bM_{\bG}$ and sub-homogeneity $\bF^{(d)}$ can be estimated accurately, the first term on the right-hand side of (\ref{yr2}) is expected to dominate $\|\widehat{\boldsymbol{\zeta}}_{m}^{(d)}\|_2$. This implies that two different clusters $d$ and $l$ are identifiable, given that $\|{\bF^{(l)}}\|_2$ has a higher order than $\|{\boldsymbol{\zeta}}_{m}^{(d)}\|_2$. As a general discussion, Theorem \ref{thm1} shows that when the sub-homogeneity from the different clusters is distinct, our clustering procedure can effectively separate the variables from the different clusters.
	
\end{remark}

\section{Simulation Studies}\label{sim}
In this section, we conduct various simulation studies to investigate the performance of our proposed CPCA compared with traditional PCA under different simulation settings.

\subsection{Simulation Settings} \label{sim setting}

We generate the data from representation (\ref{model1}):
\begin{align*}
\boldsymbol{x}_i = \sum_{k=1}^{r_c}{g}_{ik}\boldsymbol{\phi}_k + \sum_{j=1}^{J}\sum_{h=1}^{r_j}{f}^{(j)}_{ih}\boldsymbol{\gamma}^{(j)}_h + \sum_{j=1}^{J}\boldsymbol{I}^{(j)}\boldsymbol{\epsilon}^{(j)}_{i}, \quad i=1,\ldots, n, \quad j=1,\ldots, J.
\end{align*}
First, we generate $\boldsymbol{\Phi}=(\boldsymbol{\phi}_1,\ldots,\boldsymbol{\phi}_{r_c})$ as a $p\times r_c$ orthonormal matrix. Second, a $p_j\times r_j$ orthonormal matrix $\boldsymbol{\Psi}^{(j)}=(\boldsymbol{\psi}^{(j)}_1,\ldots,\boldsymbol{\psi}^{(j)}_{r_j})$ is generated randomly and independently across different clusters for $j=1,\ldots,J$ and let $\boldsymbol{\Gamma}^{(j)}_h=(\boldsymbol{0}^{(1)},\ldots,\boldsymbol{\Psi}^{(j)}_h,\ldots,\boldsymbol{0}^{(J)})$. We then generate $g_{ik}, i=1,\ldots, n, k=1,\ldots, r_c$ from $\sqrt{\delta_k}W_{ik}$, where $W_{ik}$ are i.i.d. standard normal random variables and the eigenvalues $\delta_h$ are defined as $\delta_1>\ldots>\delta_{r_c}>0$. Similarly, we generate $f^{(j)}_{ih}, i=1,\ldots, n, h=1,\ldots, r_j, j=1,\ldots, J$ from $\sqrt{\lambda^{(j)}_h} Z^{(j)}_{ih}$, where $Z^{(j)}_{hi}$ are again i.i.d. standard normal random variables and the eigenvalues $\lambda^{(j)}_h$ are defined as $\lambda^{(j)}_1>\ldots>\lambda^{(j)}_{r_j}>0, j=1,\ldots, J$. Lastly, $\boldsymbol\epsilon^{(j)}_{i}$ is a $p_j\times1$ vector consisting of $p_j$ i.i.d. normal random variables with mean $0$ and variance ${\sigma^{(j)}}^2$. Data from each cluster are generated according to the above setting and then combined to obtain $\boldsymbol{X}$.

In our simulation study, we consider four different settings, which are outlined below.
\begin{itemize}
	\item \textsc{Example} 1 ($n=50,\ p=100$). In this example, we consider the number of common components $r_c=3$ and generate $\delta_k$, $k=1,\ldots,r_c$, from $\mathcal{N}(125,5)$. We further simulate $W_{ik}$, $k=1,2,\ldots,r_c$, $i=1,2,\ldots, 2n$, from a standard normal distribution independently so that $\boldsymbol{g}_i$ can be constructed.  
	
	In terms of the cluster-specific components, we set $J=5$, and for each of the five clusters, we consider $r_j\equiv2$ and generate $\lambda_h^{(j)}\sim \mathcal{N}(\theta_{j,h},1)$, where $\theta_{j,h}$ represents the mean level of each component's variance in each cluster. In specific, we set $\theta_{1,1} = \theta_{1,2} = 25$ for the first cluster, $\theta_{5,1} = \theta_{5,2} = 5$ for the last cluster, while $\theta_{j,1} = 25$ and $\theta_{j,2} = 5$ for $j=2,3,4$, i.e., the rest three clusters.
	That is, each of the five clusters has the same number of cluster-specific components, while the components from different clusters are allowed to have different levels of variances. Among those five clusters, the components in the first cluster have the largest variances, those in the last cluster have the smallest variances, while the two components in the rest clusters are designed to have different levels of variances.
	Then, by setting $p_j\equiv20$ for each cluster, we end up with a total number of variables $p=100$. According to (\ref{model1}), $Z^{(j)}_{ih}$ for $h=1,2$ and $\boldsymbol\epsilon^{(j)}_{i}$ are simulated independently for each cluster $j$, with $\sigma^{(j)}=0.5$ for $j=1,2,...,5$ so that $2n$ observations are constructed based on (\ref{model1}). Then, the first $n$ observations are served as the training sample $\boldsymbol{X}_{train}=(\boldsymbol{x}_{1},\ldots,\boldsymbol{x}_{n})$ and the last $n$ observations are designated as the testing sample $\boldsymbol{X}_{test}=(\boldsymbol{x}_{n+1},\ldots,\boldsymbol{x}_{2n})$. In this example, $n=50$.

	\item \textsc{Example} 2 ($n=30,\ p=100$). This example is identical to Example 1, except that the sample size is decreased from 50 to 30. We will show that our CPCA is more reliable than other competing methods when the sample size is relatively small.
	
	\item \textsc{Example} 3 ($n=50,\ p=200$). In this example, we consider data consisting more clusters, that is, we consider $J=10$ clusters compared with five in Example 1, leading to a number of variables $p=200$ given $p_j\equiv20$. We maintain $r_j\equiv2$ but double the number of clusters with exactly the same setting of $\lambda_h^{(j)}\sim \mathcal{N}(\theta_{j,h},1)$. In other words, components in the first two clusters have the largest variances, while those in the last two clusters have the smallest variances. And similarly, the rest six clusters have two levels of variances on their components as described in Example 1. 
	Accordingly, $\boldsymbol\epsilon^{(j)}_{i}$ is simulated independently for each of the 10 clusters, with $\sigma^{(j)}=0.5$ for $j=1,\ldots,10$. The settings for the common components remain the same as those in Example 1.

	\item \textsc{Example} 4 ($n=30,\ p=100$). This example is identical to Example 2, but without the common effect, which is a special case of our proposed data representation as in (\ref{case 2}). Correspondingly, the response is generated as:
	\begin{align}
	{y}_i = \sum_{j=1}^{J}{\boldsymbol{f}_i^{(j)\top}}\boldsymbol{\beta}_j + {e}_{i}, \quad i=1,\ldots, n, \quad j=1,\ldots, J.
	\label{pcr example 4}
	\end{align}
	In this example, we do not estimate the common effect, and the clustering method is directly applied to the data. Therefore, the purpose of considering this example is to demonstrate that our proposed LOO-PCR clustering (Iterative Step (b) and (c) in CPCA), served as a clustering method itself, can outperform the hierarchical clustering.
\end{itemize}

For all four examples, we repeat the simulation procedure 100 times and investigate the clustering and recovering accuracy of the following methods:
\begin{itemize}
	\item PCA: The classical PCA method is performed directly on the whole data set without clustering variables. In addition, clustering accuracy is not considered for this method. The number of principal components is determined according to the ratio-based estimator (\ref{rratio}) in Remark \ref{remark1}.
	
	\item CPCA\_I: This procedure is simply CPCA, but without the Iterative Step. First, the complement $\boldsymbol{X}^c_0=\boldsymbol{X}-	\boldsymbol{G}_0\boldsymbol{\Phi}^\top_0$ is found, and then clusters $C^{(j)}_0$ are obtained using hierarchical clustering for $\boldsymbol{X}^c_0$. Variables of $\boldsymbol{X}^c_0$ are clustered, and the cluster-specific components $\boldsymbol{F}_0^{(j)}$ are finally obtained using PCA on each cluster. 
	
	
	\item CPCA\_F: This is exactly the estimation procedure described in CPCA. Different to CPCA\_I, the clusters and components are estimated iteratively. 
	
\end{itemize}

The main purposes of this comparison are to demonstrate that: 1) traditional PCA fails to capture the cluster-specific components while our proposed CPCA can successfully identify and capture the cluster-specific components; 2) our proposed iterative estimation of the clusters and components outperforms the initial estimation.


In this study, the recovering accuracy is measured by the mean squared recovering error (MSRE). In this study, the MSRE is defined as:
\begin{align}
MSRE=\frac{1}{np}\left\|\widehat{\boldsymbol{X}}_{test}-\boldsymbol{X}_{test}\right\|^2_F,
\label{MSRE}
\end{align}
where $\widehat{\boldsymbol{X}}_{test}$ is the recovered testing sample computed using the testing sample and the eigenvectors ($\boldsymbol{\Phi}, \boldsymbol{\Lambda}^{(j)}$) estimated from the training sample $\boldsymbol{X}_{train}$. We prefer a lower MSRE because it indicates a better recovering of the data.

\subsection{Simulation Results} \label{sim result}
\begin{figure}[!htbp]
	\centering
	\includegraphics[scale=0.4]{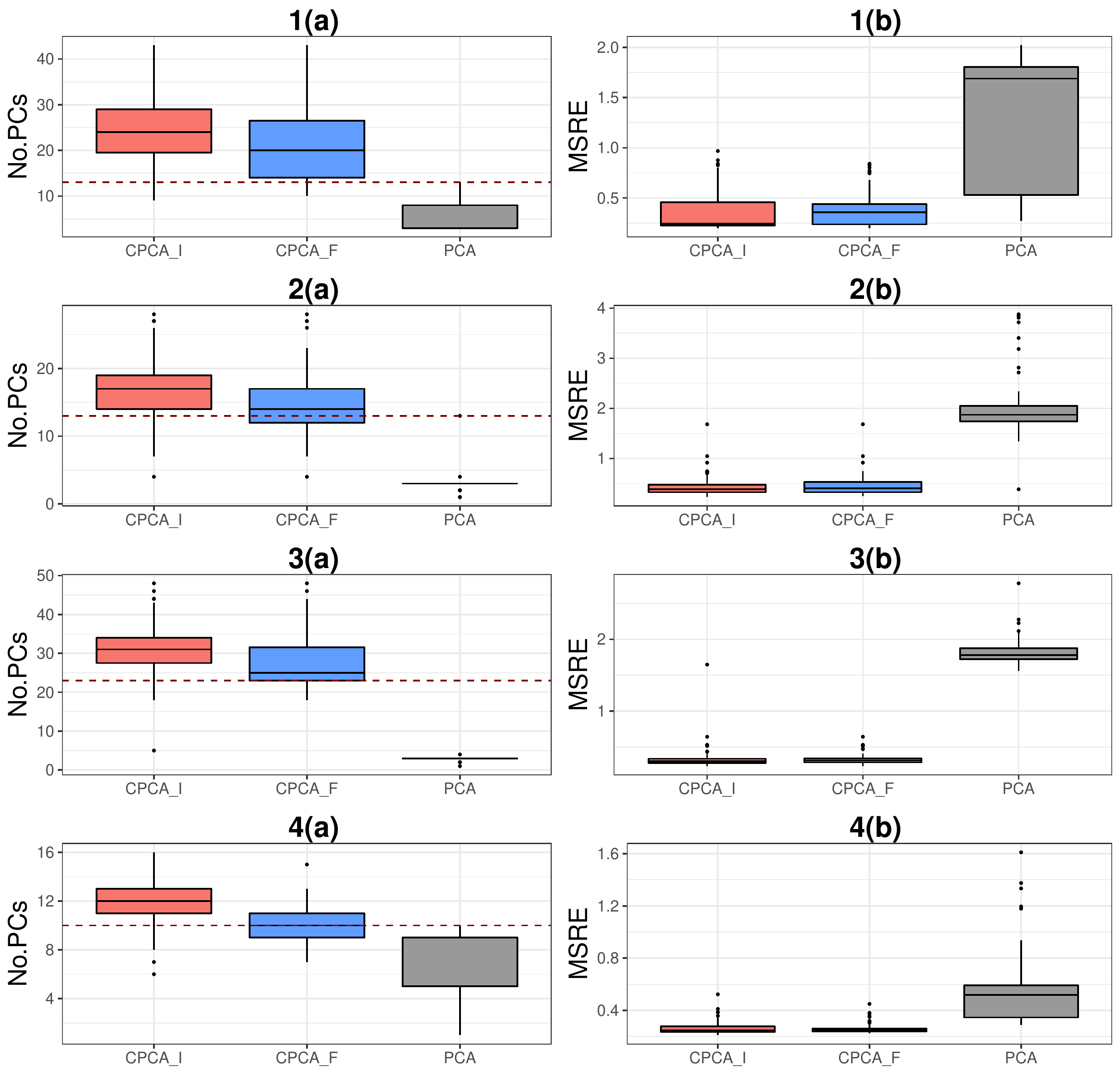}
	\caption{Boxplots of the following three measurements based on 100 simulations from Example 1 to 4: (a) No.PCs and (b) MSRE.}
	\label{fig:simulation} 
\end{figure}

The boxplots of the total number of principal components selected (No.PCs) and MSRE for Example 1,2 and 3 are presented in Figure \ref{fig:simulation} and Table \ref{simulation results}. As shown, CPCA\_I (blue boxes) achieves a consistently higher number of principal components than its iterative counterpart (red box) and the true number of principal components (dotted line), which implies the advantages of using the iterative estimation. In the initial clustering, the common effects are estimated without partitioning the variables, while the iterative clustering partitions the variables in the previous step to estimate the common components, resulting in a more accurate estimation. This is also stated in Remark \ref{remark2}. 

From the recovering accuracy perspective, it is worth noting from Figure \ref{fig:simulation} and Table \ref{simulation results} that PCA always performs poorly under this cluster structure of the data, while both CPCA-based methods demonstrate superior performance in recovering because they achieve lower MSRE than PCA. Although both CPCA\_I and CPCA\_F can achieve similar MSRE, the numbers of PCs considered by CPCA\_I are generally greater than those identified by CPCA\_F, which also indicates the importance of the iterative partitions.  
In general, CPCA\_F results in more accurate partitions and produces more reliable estimations of the cluster-specific components, especially for those with small variations, which contribute the most to predicting the response according to our simulation settings. This further confirms the advantages of using our proposed method. 

The results of our proposed CPCA\_I and CPCA\_F in Example 2 and 3 are similar, while when the sample size is smaller ($n=30$) or the number of variables and clusters are larger ($p=200$), the performance of traditional PCA is adversely affected. This is not surprising because traditional PCA estimates principal components from the sample covariance matrices which are less accurate due to the so-called `Curse of dimensionality' and only the common components can be identified. However, our proposed CPCA is superior in estimating PCs and recovering the original data as the sub-homogeneity are correctly estimated. 


Lastly, we investigate Example 4, which does not take into account the common effect. In this example, CPCA\_I and CPCA\_F simply reduce to performing PCA on each cluster of variables based on hierarchical clustering and LOO-PCR clustering, respectively. From the last row of Figure \ref{fig:simulation} and the last column of Table \ref{simulation results}, we observe that applying PCA after clustering the variables demonstrates better performance than applying PCA directly to the data. Compared with CPCA\_I, CPCA\_F achieves a better estimate of the number of PCs, implying that our proposed LOO-PCR clustering outperforms hierarchical clustering.

\renewcommand\arraystretch{1.1}
\begin{table}[!htbp] \centering 
	\caption{Averages (standard errors) of total number of principal components selected (No.PCs) and MSPE for Example 1, 2, 3, and 4.}
	\label{simulation results} 
	\begin{tabular}{@{\extracolsep{5pt}} llcccc} 
		\\[-1.8ex]\hline 
		\hline \\[-1.8ex] 
		& Method      & Example 1 & Example 2 & Example 3 & Example 4 \\ \hline
		No.PCs & CPCA\_I & 24.36 & 16.90 & 30.96 & 12.22 \\ 
		&  & (6.16) & (5.14) & (6.53) & (1.88) \\ 
		& CPCA\_F & 21.22 & 14.64 & 27.78 & 10.12 \\ 
		&  & (7.70) & (4.87) & (6.89) & (1.15) \\ 
		& PCA &  5.65 &  3.00 &  2.95 &  6.54 \\ 
		&  & (4.04) & (1.32) & (0.31) & (2.56) \\ 
		MSRE & CPCA\_I &  0.37 &  0.45 &  0.34 &  0.26 \\ 
		&  & (0.20) & (0.22) & (0.16) & (0.05) \\ 
		& CPCA\_F &  0.38 &  0.47 &  0.33 &  0.26 \\ 
		&  & (0.17) & (0.22) & (0.07) & (0.04) \\ 
		& PCA &  1.32 &  2.00 &  1.82 &  0.54 \\ 
		&  & (0.67) & (0.60) & (0.17) & (0.24) \\ 
		\hline \\[-1.8ex] 
	\end{tabular} 
\end{table} 

\section{Applications of CPCA}\label{application}
\subsection{Principal component regression}\label{application:pcr}

As aforementioned, one of the most important uses of the principal components is PCR. We consider a PCR model with an univariate response $y_i$ for $i^{th}$ observation as:
\begin{align}
{y}_i = \boldsymbol{g}^\top_i \boldsymbol{\alpha}+ \sum_{j=1}^{J}{\boldsymbol{f}^{(j)}}^\top_i\boldsymbol{\beta}_j + {e}_{i}, \quad i=1,\ldots, n, \quad j=1,\ldots, J.
\label{pcr}
\end{align}
where $\boldsymbol{\alpha}$ is a $r_c\times1$ vector representing the regression coefficients of the common components $\boldsymbol{g}_i$, $\boldsymbol{\beta}_j$ is a $r_j\times1$ vector denoting the regression coefficients of $j^{th}$ cluster-specific components $\boldsymbol{f}^{(j)}_i$ and ${e}_{i}$ is simply the error term with mean $0$ and variance $\theta^2$. 

Therefore, after the the common components $\boldsymbol{g}_i$ and cluster-specific components $\boldsymbol{f}^{(j)}_i$ are estimated via CPCA, we can fit a PCR to predict the response. However, in many situations, only a few groups are useful in predicting the response. To investigate which clusters of variable have an impact on predicting the response, we utilize the group lasso \citep{yuan2006model} using the components produced by CPCA as the coviarates to estimate the regression coefficients as in (\ref{pcr}), 
\begin{align}  
\left(\widehat{\boldsymbol{\alpha}},\widehat{\boldsymbol{\beta}}\right)=\argmin_{\boldsymbol{\alpha},\boldsymbol{\beta}}\frac{1}{2n}\sum_{i=1}^{n} \left({y}_i-\boldsymbol{g}^\top_i \boldsymbol{\alpha}- \sum_{j=1}^{J}{\boldsymbol{f}_i^{(j)\top}}\boldsymbol{\beta}_j\right)^2+\lambda\left(\left\|\boldsymbol{\alpha}\right\|_2+\sum_{j=1}^{J}\left\|{\boldsymbol{\beta}^{(j)}}\right\|_2\right),
\label{group mlr}
\end{align}
where $\lambda$ is the tuning parameter that controls the sparsity of the regression coefficients. Using this group lasso penalty, $\boldsymbol{\beta}_j$ for some $j$ will be shrunk to zero exactly so that we can identify which clusters of variables are important in predicting the response.

To better demonstrate the performance of our method, we conduct similar simulation studies as in Section \ref{sim}. We consider the following three examples:

\begin{itemize}
	\item \textsc{Example} 1 ($n=50,p=100$). We generate $\boldsymbol{X}$ using the same settings as Example 1 in Section \ref{sim setting}. Then, we simulate the response $y_i$ according to (\ref{pcr}) by setting regression coefficients $\boldsymbol{\alpha}=(1,1,1)$, $\boldsymbol{\beta}=(\boldsymbol{\beta}_1,\ldots,\boldsymbol{\beta}_{10})=(0, 0,\ldots,0,25,25)$ and standard deviation $\theta=1$. That is, only the common effect and the cluster-specific effects for the last cluster are important in predicting the response. This regression setting is interesting because the cluster-specific components with smaller variance (e.g. the last cluster) are highly likely to be omitted or estimated poorly in traditional PCA, but they can sometimes be very important in predicting the response.

	\item \textsc{Example} 2 ($n=30,p=100$). This example is identical to Example 1 above, except that the sample size is decreased from 50 to 30. 
	
	\item \textsc{Example} 3 ($n=50,p=200$). We generate $\boldsymbol{X}$ using the same settings as Example 3 in Section \ref{sim setting}. That is, we consider $J=10$ clusters in this example. In regard to the PCR, we set regression coefficients $\boldsymbol{\alpha}=(1,1,1)$ again and consider $\boldsymbol{\beta}=(\boldsymbol{\beta}_1,\ldots,\boldsymbol{\beta}_{20})=(0, 0,\ldots,0,25,25)$ such that only the common effect and the cluster-specific effects for the last cluster are useful and the rest of clusters are noise.
	
\end{itemize}
For all three examples, we repeat the simulation procedure 100 times and investigate prediction accuracy of the methods mentioned in Section \ref{sim setting}. The MSPE is used to determine whether the selected components can accurately predict the response:
\begin{align}
\text{MSPE}=\frac{1}{n}\sum_{i=1}^n\left(\widehat{y}_{i,test}-{y}_{i,test}\right)^2,
\label{MSPE}
\end{align}
where $\widehat{y}_{i,test}$ is computed using (\ref{pcr}) with the regression coefficients ($\boldsymbol{\gamma},\boldsymbol{\beta}$) and the eigenvectors $\left(\boldsymbol{\Phi}, \boldsymbol{\Lambda}^{(j)}\right)$ estimated from the training sample.

\begin{figure}[!htbp]
	\centering
	\includegraphics[scale=0.43]{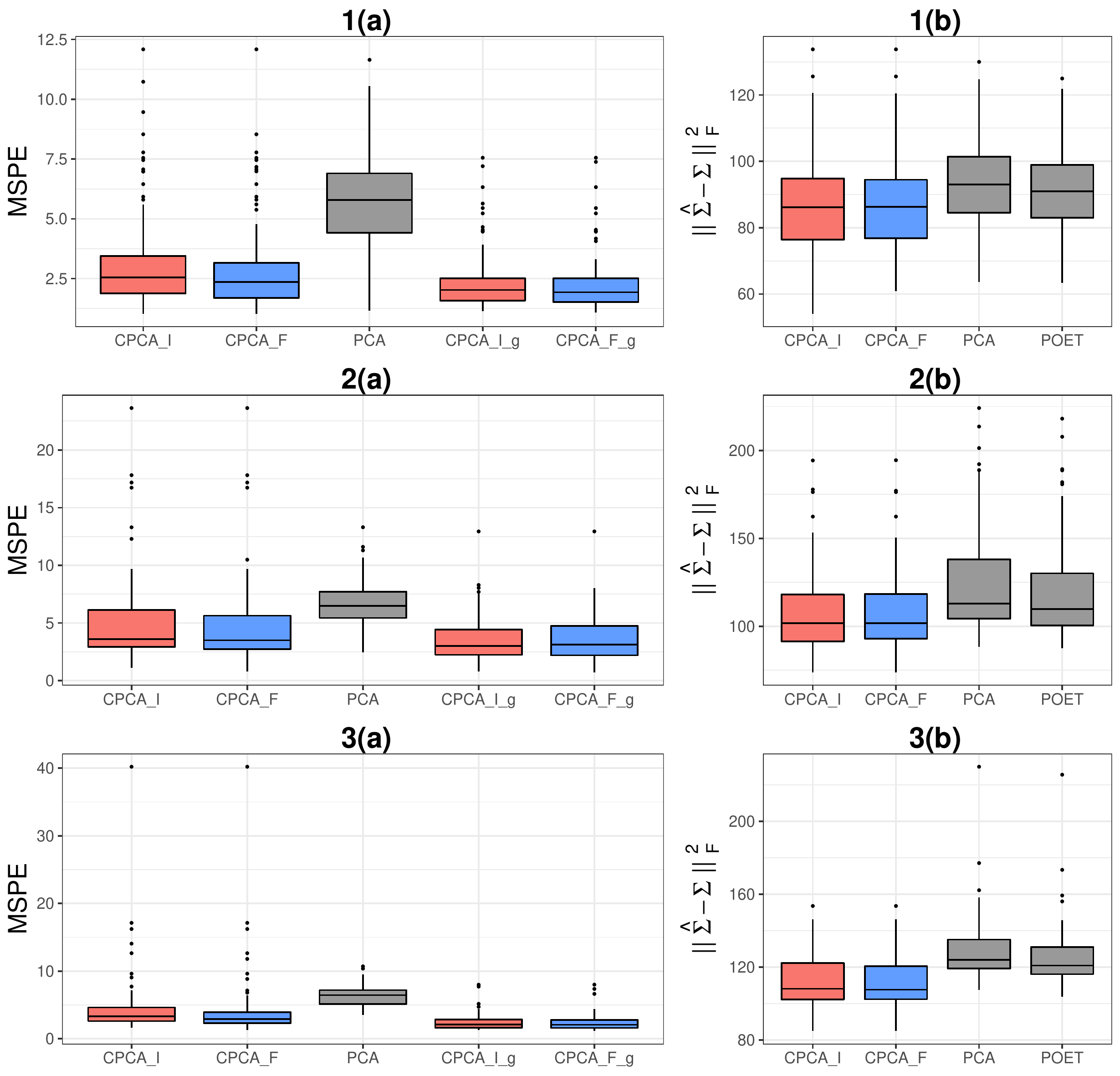}
	\caption{Boxplots of the following two measurements based on 100 simulations from Example 1 to 3: (a) MSPE and (b) $\left\|\widehat{\boldsymbol{\Sigma}}-\boldsymbol{\Sigma}\right\|^2_F$ .}
	\label{fig:application} 
\end{figure}

\renewcommand\arraystretch{1.1}
\begin{table}[htb] \centering 
	\caption{Averages (standard errors) of  MSPE and $\left\|\widehat{\Sigma} - \Sigma\right\|_\text{F}^2$  for Example 1, 2, and 3.}
	\label{application results} 
	\begin{tabular}{@{\extracolsep{5pt}} llccc} 
		\\[-1.8ex]\hline 
		\hline \\[-1.8ex]
		& Method  & Example 1 & Example 2 & Example 3 \\ \hline
		MSPE & CPCA\_I &  3.27 &   5.30 &   4.65 \\ 
		&  & (2.14) & (4.21) & (4.92) \\ 
		& CPCA\_F &  2.95 &   4.90 &   4.22 \\ 
		&  & (1.92) & (4.12) & (4.94) \\ 
		& PCA &  5.55 &   6.67 &   6.32 \\ 
		&  & (2.37) & (2.03) & (1.43) \\ 
		& CPCA\_I\_g &  2.38 &   3.66 &   2.43 \\ 
		&  & (1.27) & (2.13) & (1.18) \\ 
		& CPCA\_F\_g &  2.30 &   3.61 &   2.40 \\ 
		&  & (1.23) & (2.10) & (1.19) \\ 
		$\|\widehat{\Sigma} - \Sigma\|_\text{F}^2$ & CPCA\_I & 85.92 & 108.55 & 111.71 \\ 
		&  & (14.83) & (25.05) & (13.57) \\ 
		& CPCA\_F & 86.41 & 108.80 & 111.52 \\ 
		&  & (14.01) & (24.71) & (13.33) \\ 
		& PCA & 93.40 & 123.95 & 128.53 \\ 
		&  & (12.83) & (31.88) & (16.77) \\ 
		& POET & 91.28 & 119.71 & 125.01 \\ 
		&  & (11.87) & (30.49) & (16.46) \\ 
		\hline \\[-1.8ex] 
	\end{tabular} 
\end{table} 

The first panel of Table \ref{application results} and the first column of Figure \ref{fig:application} display mean and standard errors and the boxplot of MSPE for all three examples. Note that in this section, CPCA\_F and CPCA\_F\_g represent the ordinary least squares (OLS) and the group lasso regressed on the principal components produced by CPCA\_F, respectively. Similar notations apply to the rest of CPCA-based methods. When the traditional PCR (OLS) is utilized, we clearly see that CPCA\_F achieves much lower average MSPE and standard deviation than other methods, especially when $n$ is small or $p$ is large. Not surprisingly, the traditional PCA performs the worst because it fails to capture any of the sub-homogeneity, which is important in predicting the response under our setting. Moreover, we observe that PCR using the group lasso also outperforms those OLS counterparts, especially for CPCA\_I. The group lasso can provide substantial reductions in MSPE for CPCA\_I because it tends to select a large number of principal components, only some of which are useful. Overall, CPCA\_F\_g achieves the lowest mean and standard error of MSPE among all the methods in all three examples, because it not only accurately extracts both common components and group-specific components, but also shrinks the coefficients of those unimportant groups down to zero.

\subsection{Covariance and Precision Matrix Estimation}\label{application:cov}
Another useful application of CPCA is to estimate $\boldsymbol{\Sigma}$ as in (\ref{cov}). Using $\boldsymbol{g}_i$ and $\boldsymbol{f}^{(j)}_i$ leads to a better estimation of $\boldsymbol{\Sigma}$ compared with using $\boldsymbol{g}_i$ only, because  $\boldsymbol{g}_i$ and $\boldsymbol{f}^{(j)}_i$ can capture both the low rank and the block-diagonal representation of $\boldsymbol{\Sigma}$. Therefore, we estimate $\boldsymbol{\Sigma}$ using clusters, the common components, and the cluster specific components, along with their associated eigenvectors estimated via CPCA. In this section, we numerically illustrate the advantage of using CPCA to estimate the covariance matrix in comparison with other traditional PCA methods. Again, we conduct three simulation examples mentioned in Section \ref{application:pcr} and utilize the Euclidean distance (ED) between the estimated covariance and population covariance, $\|\widehat{\boldsymbol{\Sigma}}-\boldsymbol{\Sigma}\|^2_F$, to measure the performances of different methods. Recall $\boldsymbol{\Sigma}$ is generated from (\ref{cov}):
\begin{align*}
\boldsymbol{\Sigma}= \boldsymbol{\Phi}\text{cov}\left(\boldsymbol{g}_i \right)\boldsymbol{\Phi}^\top+ \sum_{j=1}^{J}\boldsymbol{\Gamma}^{(j)}\text{cov}(\boldsymbol{f}^{(j)}_i) {\boldsymbol{\Gamma}^{(j)}}^\top+ \sum_{j=1}^{J}\boldsymbol{I}^{(j)}{\sigma^{(j)2}}.
\end{align*}
In this section, we add another prevalent covariance estimation method POET \citep{fan2013large} as aforementioned into our comparison. 

The second panel of Table \ref{application results} and the second column of Figure \ref{fig:application} demonstrate the mean and standard errors and boxplot of 
$\|\widehat{\boldsymbol{\Sigma}}-\boldsymbol{\Sigma}\|^2_F$ for all three examples, computed using methods discussed before. From these results, we see that POET performs better than PCA but worse than CPCA\_I and CPCA\_F, indicating that the sparsity structure implemented in POET can partly capture the sub-homogeneity, while not in a very efficient way. Among these methods, both CPCA\_I and CPCA\_F can achieve the lowest mean and standard error of $\|\widehat{\boldsymbol{\Sigma}}-\boldsymbol{\Sigma}\|^2_F$ because it can best identify both homogeneity and sub-homogeneity to accurately estimate $\boldsymbol{\Sigma}$.

\section{Real Data Analysis}\label{real}
For further illustration, we analyze a stock return data set using the proposed CPCA and compare it with traditional PCA. In particular, we are interested in the performance of our proposed CPCA on clustering stock returns of companies from various industries. Moreover, we also use CPCA to obtain an estimate of the covariance matrix and use it to construct a minimum variance portfolio (MVP).
\subsection{Clustering Stock Returns}
The data are collected from the Center for Research in Security Prices (CRSP) and include the daily stock returns of 160 companies from 1st January, 2014 to 31st December, 2014, with 252 trading days. The 160 stocks are selected from eight different industries according to Fama and French’s 48-industry classification \citep{fama1997industry}, namely, Candy and Soda, Tobacco Products, Apparel, Aircraft, Shipbuilding and Railroad Equipment, Petroleum and Natural Gas, Measuring and Control Equipment, and Shipping Containers, with 20 stocks from each industry. The data for the first 126 trading days are treated as the training sample, and the rest are the testing sample. Thus, the training data have the dimensions $n=126$ and $p=160$. In this example, after removing the common effect from the data, we aim to identify the clusters that consist of companies from the same industry. This cluster structure of stock returns is also discussed in \citet{fan2013large}.

\begin{figure}[htb]
	\centering
	
	\subfloat[][]{\includegraphics[scale=0.27]{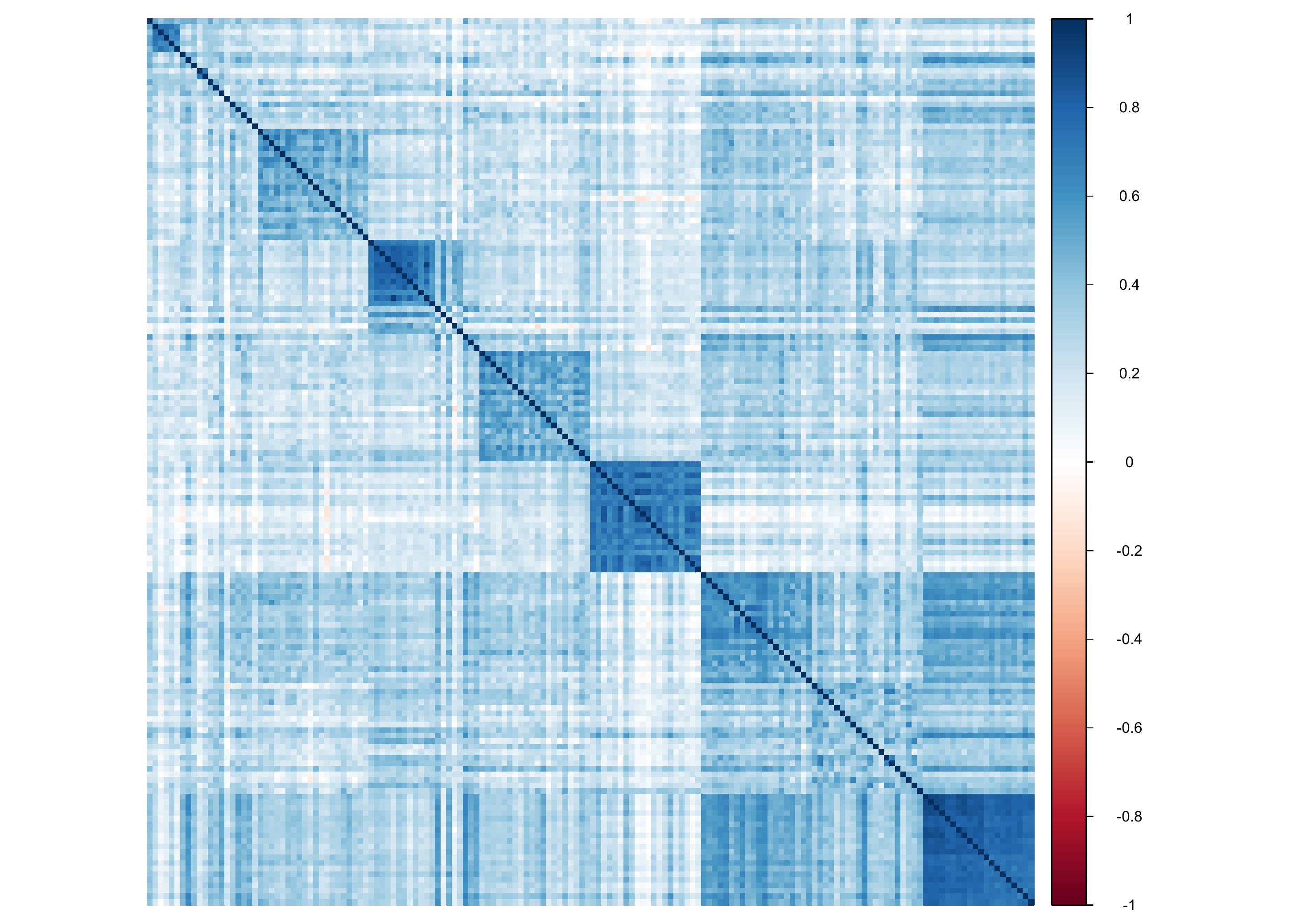}}\subfloat[][]{\includegraphics[scale=0.27]{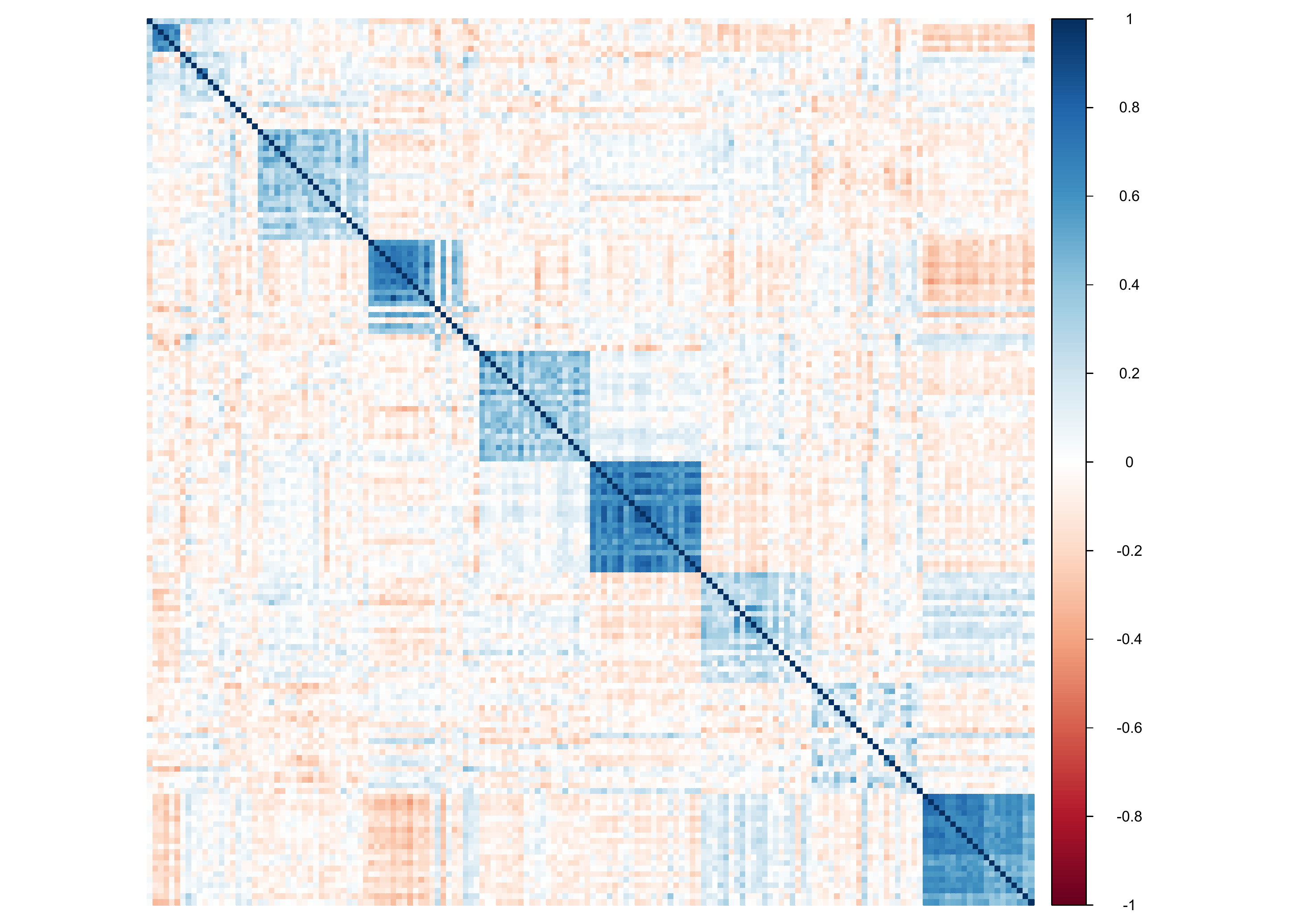}}\\  [10pt]
	
	\caption{Correlation plot for (a) the original stock data $\boldsymbol{X}$, (b) the complement of the stock data $\boldsymbol{X}^c_f=\boldsymbol{X}-\boldsymbol{G}_f\boldsymbol{\Phi}^\top_f$}
	
	\label{fig:corplotstock}
\end{figure}

\begin{figure}[htb]
	\centering
	\includegraphics[scale=0.6]{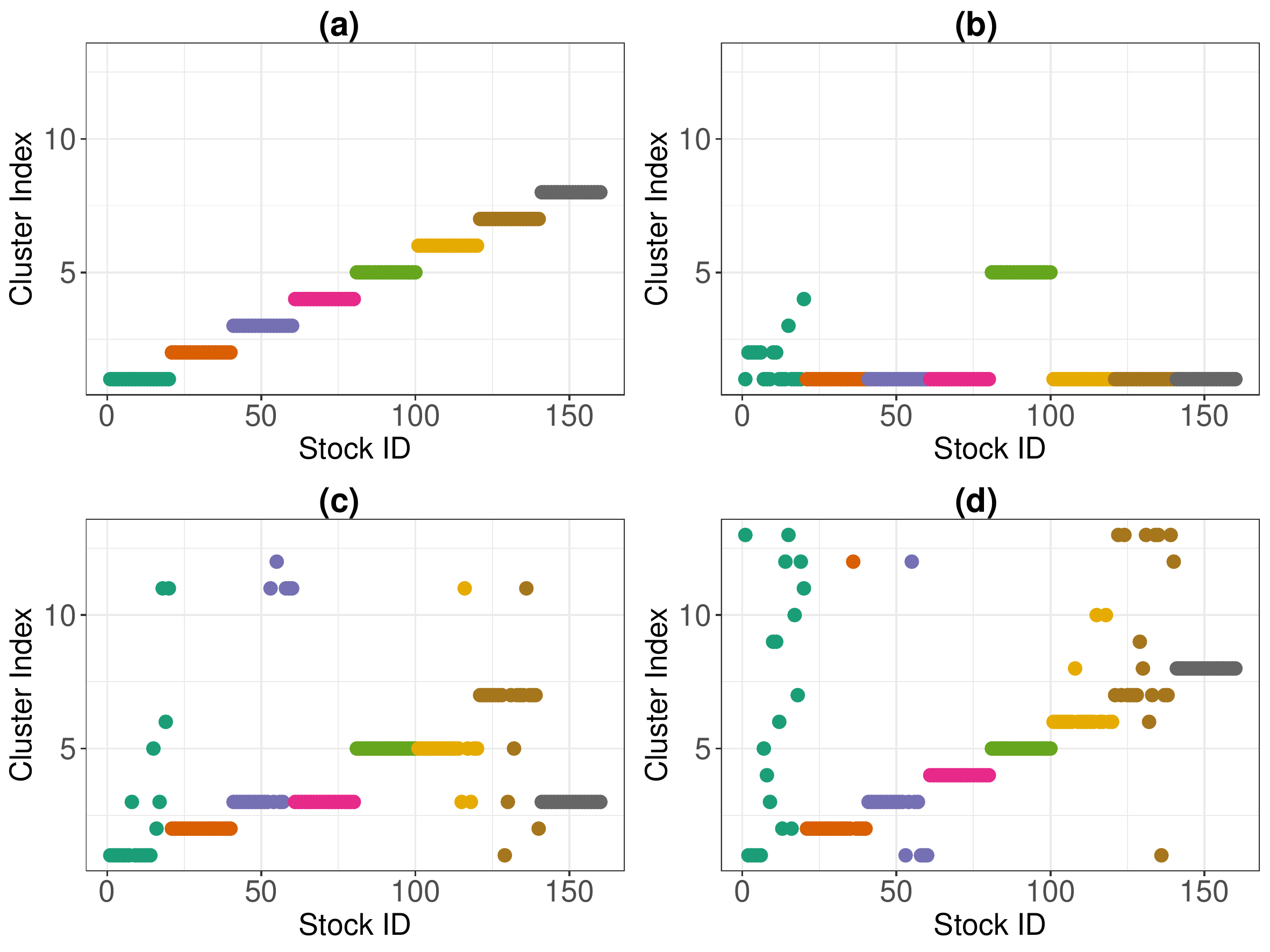}
	\caption{Cluster membership for 160 stocks based on (a) their industry, (b) hierarchical clustering of the original data, (c) initial clusters obtained from CPCA, (d) final clusters obtained from CPCA. Stocks with the same color are from the same industry.}
	\label{fig:index}
\end{figure}

Figure \ref{fig:corplotstock} (a) shows the correlation plot for the original stock data. We can observe a vague cluster structure, but the off-diagonals are clearly non-zero, implying that the stocks from the different industries selected in our study tend to be positively correlated. We apply hierarchical clustering directly to the original data and find that most stocks from different industries are clustered in the same group, as displayed in Figure \ref{fig:index} (b). Only stocks from the $5^\text{th}$ industry stand out as another cluster, because they have a relatively lower correlation with other stocks. If we consider the industries as true clusters, the hierarchical clustering of the original data results in an ARI of only $0.09$. This indicates that common components may exist and conceal the cluster structure within the data.

Next, we apply our proposed CPCA to the stock data, and one common component is determined using the ratio-based estimator (\ref{rratio}). 
Figure \ref{fig:corplotstock} (b) presents the correlation plot of the data after the common effect is removed. Comparatively, the cluster structure is more apparent. One can interpret the common effect as the market effect and the rest of the components as industry-specific effects. 
As demonstrated in Figure \ref{fig:index} (c), after the initial step of CPCA, a few clusters of stocks from the $1^\text{st},\ 2^\text{nd},\ \text{and}\ 7^\text{th}$ industries are correctly identified, while stocks from other industries (the $3^\text{rd},\ 4^\text{th},\ 5^\text{th},\ 6^\text{th},\ \text{and}\ 8^\text{th}$) cannot be correctly separated from two large clusters. The ARI after the initial step of CPCA is about $0.44$ which is a significant improvement to $0.09$. Figure \ref{fig:index} (d) depicts the final clusters after the iterative steps of CPCA where most stocks from the aforementioned two large clusters can be correctly separated according to industries with a minor cost of a few miss-classifications in the $1^\text{st}\  \text{and}\ 7^\text{th}$ cluster. The ARI is also further improved from $0.44$ to $0.72$. The final clusters mainly capture the industry information, but a few stocks are not clustered into their own industries. This is to be expected because some stocks from the same industry are not highly correlated, as shown in Figure \ref{fig:corplotstock} (e.g., some stocks from the first, third, sixth and seventh industries), which can be common in reality due to diversification of enterprises.

From the prediction perspective, we compute the MSRE for PCA, CPCA\_I, and CPCA\_F as $3.56$, $2.23$, and $1.67$ (in unit of $10^{-4}$), respectively. In other words, our CPCA can not only capture the market effect of stock returns as the homogeneity but also include some sub-homogeneities which can be interpreted as industrial effects. All of these findings firmly support that our proposed CPCA is appropriate for analyzing these stock return data.

\subsection{Constructing Minimum Variance Portfolio}
Next, we consider the problem of constructing a global minimum variance portfolio (MVP) from stock returns. The minimum variance portfolio is built by solving the following optimization problem
\begin{align*}
	\min_{\boldsymbol{w}}\ &\boldsymbol{w}^\top \boldsymbol{\widetilde{\Sigma}} \boldsymbol{w} \\
	\text{s.t. } &\boldsymbol{w}^\top \boldsymbol{1}_p = 1,
\end{align*}
where $\boldsymbol{w}$ is a $p$-dimensional vector of weights allocated on each stock, $\boldsymbol{1}_p$ is a $p$-dimensional vector of ones and $\boldsymbol{\widetilde{\Sigma}}$ is defined as an estimate of the $p$ by $p$ covariance matrix of stock returns. The analytic solution of $\boldsymbol{w}$ is
\begin{align}\label{weight}
	\boldsymbol{{w}}^{\star} = \frac{\boldsymbol{\widetilde{\Sigma}^{-1}}\boldsymbol{1}_p}{\boldsymbol{1}_p^\top\boldsymbol{\widetilde{\Sigma}^{-1}}\boldsymbol{1}_p},
\end{align}
where we assume short sale is allowed and there are no transaction costs.

Constructing the global MVP is another important application of our proposed CPCA due to its ability of correctly identifying both the homogeneity and the sub-homogeneity when estimating the covariance matrices of stock returns. The application of constructing MVP have been well studied and discussed by \citet{engle_large_2019,chen_new_2019} and \citet{wang2021nonparametric} where several different estimation approaches for the covariance matrices of stock returns are developed.
Our interest in this section is the numerical results of the out-of-sample performance of the global MVP constructed using the covariance matrices estimated by CPCA. Recall that stock returns of $160$ companies from $8$ industries are collected for $252$ trading days, we use rolling windows with the size of $110$ days (5 calendar months) to obtain the out-of-sample performance of the portfolio. In specific, for each trading day $i$, we firstly estimate $\boldsymbol{\widetilde{\Sigma}}$ by CPCA where stock returns in the most recent $110$ tradings days before day $i$ (from day $(i-110)$ to day $(i-1)$) are considered, and then compute the optimal weight $\boldsymbol{{w}}_i^{\star}$ by (\ref{weight}). Consequently, the return of the global MVP on day $i$ can be computed using $\boldsymbol{{w}}_i^{\star}$ and the return of each stock on day $i$. Similar to \citet{wang2021nonparametric}, we compute the standard deviation (STD), the information ratio (IR), which is the ratio of mean return to STD, and the Sharpe ratio (SR), which is the ratio of mean excess return (the portfolio-wide return minus the risk-free return) to STD. As discussed by \citet{engle_large_2019}, \citet{chen_new_2019} and \citet{wang2021nonparametric}, the STD should be the primary measure when comparing the MVPs constructed using different estimates of covariance matrices, while the IR and SR are also desirable but less crucial. The numerical results in Table \ref{MVP} compare the portfolios constructed with different estimates of the covariance matrices. As depicted, CPCA with both the initial and final clusters can achieve a slightly better estimate of the covariance matrix, since the portfolio corresponding to CPCA with has the smallest STD. In addition, the global MVP constructed using CPCA with the final clustering index can also achieve highest IR and SR across all methods, which is also desirable. As a result, CPCA works well when constructing the global MVP since the sub-homogeneity of industrial effects can be well estimated.

\renewcommand\arraystretch{1.1}
\begin{table}[htb] \centering 
	\caption{Out-of-sample performance of the minimum variance portfolio (MVP) constructed using the covariance matrices estimated by traditional PCA, CPCA with initial clustering index, CPCA with final clustering index and POET method.}
	\label{MVP} 
	\begin{tabular}{@{\extracolsep{5pt}} lccc} 
		\\[-1.8ex]\hline 
		\hline \\[-1.8ex]
		Method  & STD    & IR     & SR     \\ \hline
		CPCA\_I & 0.008 & 0.120 & 0.112 \\
		CPCA\_F & 0.008 & 0.151 & 0.143 \\
		PCA     & 0.010 & 0.107 & 0.099 \\
		POET    & 0.009 & 0.148 & 0.140 \\
		\hline \\[-1.8ex] 
	\end{tabular} 
\end{table} 

\section{Discussion}\label{con}
To conclude, we introduced a novel CPCA to study the homogeneity and sub-homogeneity of high-dimensional data collected from different populations, where the sub-homogeneity refers to a group-specific feature from a particular population. Our numerical simulations confirmed that traditional PCA can only extract the homogeneity from the data, whereas CPCA not only provides a more accurate estimate of the common components, but also identifies the group-specific features, even for the group with small variations. Under a PCR setting, the features extracted using CPCA can significantly outperform those selected by classical PCA in terms of prediction, especially when $n$ is small  but $p$ is large. Our real analysis of the stock return data also demonstrated that, when using CPCA, we can capture the industry information after the common component (i.e., market effect) is removed. All of these findings support the use of our proposed CPCA in dimension-reduction problems.

The applications of CPCA are not limited to producing principal components in PCR and revealing the industry structure from the stock return data. CPCA can also be applied to any other data sets in which a group structure exists but hides in the homogeneity. Further, it can be used to estimate a covariance matrix and its inverse of a large data set that exhibits a group structure. More applications of CPCA will be explored in our future work.

\bigskip
\begin{center}
	{\large\bf SUPPLEMENTARY MATERIAL}
\end{center}

\begin{description}
	
	\item[R-codes for CPCA:] R-codes including R functions to perform the iterative complement-clustering principal component analysis (CPCA) described in the article. Also included are the R-codes for simulation studies and real analysis of the stock return data shown in the article. (.R files)
	
	\item[Data of stock returns:] Data set of daily returns of stocks from different industries in the US (2014). (.rds file)

\end{description}

\bibliographystyle{agsm}
\bibliography{reference}

\newpage
\begin{appendices}
	\section{Estimation of $\bG_1$ and $\bPhi$ in CPCA Iterative Step (a)}\label{append A}
	
	In this part, we describe the estimation of the principal components, along with their eigenvectors, combined from each cluster in Iterative Step (a) of Algorithm 1.
	
	First, we perform PCA in cluster $j$ by:
	\begin{align*}
	\bX^{(j)} =  \bPsi^{(j)}\bPi^{(j)\top} + \bU^{(j)},
	\end{align*}
	where $\bPsi^{(j)}$ is an $n\times r_j$ matrix of principal components for variables in cluster $j$ and $\bPi^{(j)}$ is a $p_j\times r_j$ matrix in which each column represents an eigenvector of $\bX^{(j)\top}\bX^{(j)}$. Here, recall that $p_j$ denotes the number of variables in cluster $j$, and $r_j$ is the number of principal components in cluster $j$. Then, we can combine the principal components $\bPsi^{(j)}$ from each cluster as $\bPsi = (\bPsi^{(1)},...,\bPsi^{(J_0)})$ and perform a further step of PCA on $\bPsi$ to obtain:
	\begin{align*}
	\bPsi =  \bG_1\bEtat + \bV,
	\end{align*}
	where we assume the first $r_c$ principal components can be used to summarize the common effects among all clusters. Then, $\bG_1$ is a $n\times r$ matrix of principal components of $\bPsi$, and $\bEta$ is a $(\sum_{j=1}^{J_0} r_j) \times r_c$ matrix in which each column represents an eigenvector of $\bPsi^\top\bPsi $.
	Lastly, we can find the complement for cluster $j$ as:
	\begin{align*}
	\bX_1^{(j)c} = \bX^{(j)} - \bG_1\bEtat\bPi^{(j)\star\top},
	\end{align*}
	where $\bPi^{(j)\star} = (\boldsymbol{0}^{(1)},\boldsymbol{0}^{(2)},...,\bPi^{(j)},...,\boldsymbol{0}^{(J_0)})$ is a $p_j \times (\sum_{j=1}^{J_0} r_j)$ matrix in which $\bPi^{(j)}$ is a $p_j\times r_j$ matrix, as we described earlier, and $\boldsymbol{0}^{(l)}$ is a $p_j \times r_l$ zero matrix for $l=1,2,...,J_0;\ l\neq j$. Once we obtain the complement from each cluster, we can combine them and present the total complement using:
	\begin{align*}
	\bX_1^{c} = \bX - \bG_1\bEtat\bPit,
	\end{align*}
	where $\bPi = (\bPi^{(1)\star\top},\bPi^{(2)\star\top},...,\bPi^{(j)\star\top},...,\bPi^{(J_0)\star\top})^\top$ is a $p \times (\sum_{j=1}^{J_0} r_j) $ block-diagonal matrix. Hence, we can finally define $\bPhi=\bPi\bEta$ as the corresponding eigenvectors for $\bG_1$. It is also easy to show that $\bPhi^\top\bPhi=I_p$. 
	
	\section{Proof of Theorem \ref{thm1}}\label{append B}
	The matrix form of (\ref{y1}) is:
	\begin{eqnarray}
	\bu_{m}=\bF^{(d)}\boldsymbol{\beta}+\boldsymbol{\zeta}^{(d)}_{m}, 
	\end{eqnarray}
	where $\bu_{m}=\left(u_{m1}^{(l)}, u_{m2}^{(l)}, \ldots, u_{mn}^{(l)}\right)^\top$ and $\boldsymbol{\beta}=(\beta_1, \beta_2, \ldots, \beta_{r_d})$. \\
	Hence, the least-squares estimation of the residual $\boldsymbol{\zeta}^{(d)}_{m}$ is:
	\begin{eqnarray}\label{yr3}
	\widetilde{\boldsymbol{\zeta}}^{(d)}_{m}
	=\bu_{m}-\bF^{(d)}\left(\bF^{(d)\top}\bF^{(d)}\right)^{-1}\bF^{(d)\top}\bu_{m}=:\bM_{\bF^{(d)}}\bu_{m}. 
	\end{eqnarray}
	
	It should be noted that $\bu_{m}$ and $\bF^{(d)}$ are not observed. In terms of our proposed clustering method, we estimate $\bu_{m}$ via PCA, such that:
	\begin{eqnarray}\label{yr4}
	\widehat{\bu}_{m}=\bM_{\widehat{\bG}}\bx_{m}, 
	\end{eqnarray}
	where $\bM_{\widehat{\bG}}=\bI_n-\widehat{\bG}(\widehat{\bG}^{\top}\widehat{\bG})^{-1}\widehat{\bG}^{\top}$, with $\widehat{\bG}$ being an estimator for $\bG$, as described in CPCA.\\
	For $\bF^{(d)}$, we also perform PCA on cluster $d$ and denote the estimator by $\widehat{\bF}^{(d)}$. Based on (\ref{yr3}) and (\ref{yr4}), our estimator of $\boldsymbol{\zeta}^{(d)}_{m}$ can be written as:
	\begin{eqnarray}\label{yr5}
	\widehat{\boldsymbol{\zeta}}^{(d)}_{m}
	&=&\bM_{\widehat{\bF}^{(d)}}\widehat{\bu}_{m}
	=\bM_{\widehat{\bF}^{(d)}}\bM_{\widehat{\bG}}\bx_{m}\nonumber \\
	&=&\bM_{\widehat{\bF}^{(d)}}\left(\bM_{\widehat{\bG}}-\bM_{\bG}\right)\bx_{m}
	+\bM_{\widehat{\bF}^{(d)}}\bM_{\bG}\bx_{m}\nonumber \\
	&=&\left(\bM_{\widehat{\bF}^{(d)}}-\bM_{\bF^{(d)}}\right)\left(\bM_{\widehat{\bG}}-\bM_{\bG}\right)\bx_{m}
	+\bM_{\bF^{(d)}}\left(\bM_{\widehat{\bG}}-\bM_{\bG}\right)\bx_{m}\nonumber \\
	&&+\left(\bM_{\widehat{\bF}^{(d)}}-\bM_{\bF^{(d)}}\right)\bM_{\bG}\bx_{m}
	+\bM_{\bF^{(d)}}\bM_{\bG}\bx_{m}. 
	\end{eqnarray}
	To define $\|\bM_{\widehat{\bF}^{(d)}}-\bM_{\bF^{(d)}}\|_2=O_p(\alpha_{np_d})$, $\|\bM_{\widehat{\bG}}-\bM_{\bG}\|_2=O_p(\gamma_{nJ})$, we obtain the result in this theorem.

\end{appendices}

\end{document}